\documentclass[epj,nopacs]{svjour}
\usepackage{amsmath,amssymb,amsfonts,latexsym,dcolumn,graphics}
\begin{document}
\title{Quantum electrodynamics of inhomogeneous anisotropic media}
\author{Adri\'an E. Rubio L\'opez \and Fernando C. Lombardo
}                     
%
%
\institute{Departamento de F\'\i sica {\it Juan Jos\'e
Giambiagi}, FCEyN UBA and IFIBA CONICET-UBA, Facultad de Ciencias Exactas y Naturales,
Ciudad Universitaria, Pabell\' on I, 1428 Buenos Aires, Argentina \email{arubio@df.uba.ar}}
\date{Received: date / Revised version: date}
%
\abstract{
In this work we calculate the closed time path (CTP) generating functional for the electromagnetic (EM) field interacting with inhomogeneous anisotropic matter.  For this purpose, we first find a general expression for the electromagnetic field's influence action from the interaction of the field with a composite environment consisting in the quantum polarization degrees of freedom in each point of space, at arbitrary temperatures, connected to thermal baths. Then, we evaluate the generating functional for the gauge field, in the temporal gauge, by implementing the Faddeev-Popov procedure. Finally, through the point-splitting technique, we calculate closed expressions for the energy, the Poynting vector and the Maxwell tensor in terms of the Hadamard propagator. We  show that all the quantities have contributions from the field's initial conditions and also from the matter degrees of freedom. Throughout the whole work we discuss and give insights about how the gauge invariance must be treated in the formalism when the EM field is interacting with inhomogeneous anisotropic matter. We study the electrodynamics in the temporal gauge, obtaining the EM field's equation and a residual condition. Finally analyze the case of the EM field in bulk material and also discuss several general implications of our results in relation with the Casimir physics in a nonequilibrium scenario.
} 
\maketitle
%
\section{Introduction}

The main subject of this paper is to develop a CTP-integral formulation of nonequilibrium quantum electrodynamics in an inhomogeneous, anisotropic real medium. The CTP-method has been used in quantum field theory as a tool to nonequilibrium descriptions of dynamical problems, where dissipative effects arise at the macroscopic level after coarse graining the detailed information in one or more subsystems, by tracing out those degrees of freedom. In fact, this approach presents a combination of both quantum field theory and nonequilibrium statistical mechanics. The former is needed by the quantum electromagnetic (EM) field and the latter for treating processes involving quantum dissipation and noises. As far as we know, this complete formulation has not been applied previously to the quantum theory of the electromagnetic field in interaction with matter capable to dissipate and absorb energy to and from an external thermal environment, at a microscopic level. The general problem is of both basic and practical importance. The polarization properties of real media change the properties of the EM field dramatically in comparison with the situation in vacuum, especially in the case of a dispersive and absorbing dielectric. Therefore, Casimir effect appears as one macroscopic manifestation of these effects, for which it is important to have a complete theoretical formulation to contrast first principle based models with experiments.

The very well known Lifshitz formula \cite{Lifshitz} describes the forces between dielectrics in terms of their macroscopic EM properties, in thermal equilibrium. Its original derivation is not based on a first principle quantum framework, but on a macroscopic approach, starting from stochastic Maxwell
equations and using thermodynamical properties for the stochastic fields. The connection between this approach and one based on a fully quantized model including lossy dielectrics is not completely clear \cite{Barton2010,Philbin2010,daRosaetal}.

When dealing with a composite system, in which there are noise, fluctuations, and also dissipative effects between different parts of the full system (mirrors, vacuum field and environment), the theory of open quantum systems \cite{BreuerPett} is the most appropriate framework to clarify the role of
these effects in Casimir physics. Indeed, in this description, dissipation and noise appear in the effective theory of the relevant degrees of freedom (the EM field) after integration of the matter and other environmental degrees of freedom.

The quantization at the steady situation can be performed starting from the macroscopic Maxwell equations, and including noise terms to account for absorption \cite{Buhmann2007} (one can also to couple the EM field to an external reservoir \cite{Philbin2010}, following the standard route to include dissipation). Regarding microscopic models, the fully canonical quantization of the EM field in dispersive and lossy dielectrics has been performed by Huttner and Barnett (HB) \cite{HB}. In their model, the EM field is coupled to matter (the polarization field), and the matter is coupled to a reservoir that is included into the model to describe the losses. In the context of the theory of quantum open systems, one can think the HB model as a composite system in which the relevant degrees of freedom belong to two subsystems (the EM field and the matter), and the matter degrees of freedom are in turn coupled to an environment (the thermal reservoir). The indirect coupling between the electromagnetic field and the thermal reservoir is responsible for the losses.

In Ref. \cite{LombiMazziRL}, we have followed an equilibrium canonical quantization program similar to that of Ref.\cite{Dorota1992}, generalizing it by considering a general and well defined open quantum system. Recently, in Ref.\cite{RubLoLom}, we have considered two simplified models analogous to the one of HB, both assuming that the dielectric atoms in the slabs are quantum Brownian particles, and that they were subjected to fluctuations (noise) and dissipation, due to the coupling to an external thermal environment, also generalizing the constant dissipation model of Ref.\cite{LombiMazziRL}. Indeed, after integration of the environmental degrees of freedom, it was possible to obtain the dissipation and noise kernels that modify the unitary equation of motion of the dielectric atoms. Using the Schwinger-Keldysh formalism (or closed time path (CTP) - in-in - formalism) \cite{Schwinger,Keldish}, we have studied the time evolution of the expectation value of the energy-momentum tensor of a scalar field in the presence of real materials. The work included a fully nonequilibrium scenario done for two different couplings between the scalar field and the polarization degrees of freedom of matter. There it was shown that two contributions always take place in the transient evolution of the energy-momentum tensor: one associated with the material, and other one only related to the field. Therefore, we have shown that the material always contribute unless is non-dissipative. Conversely, the proper field contribution vanishes unless the material is non-dissipative or, moreover, at least for the $1+1$ case, if there are regions without material. We finally concluded that any steady quantization scheme in $1+1$ dimensions must consider both contributions and, on the other hand, we argued why these results are physically expected from a dynamical point of view, and also could be valid for higher dimensions based on the expected continuity between the non-dissipative and real material cases \cite{RubLoLom}.

In this work, we will extend Ref. \cite{RubLoLom} in order to calculate the CTP generating functional for the EM field (abelian gauge field) interacting with inhomogeneous and anisotropic matter through the open system framework. For this purpose, we will firstly calculate a general expression for the EM field's influence action from the interaction of the field with a composite environment consisting in quantum polarization degrees of freedom in each point of space (having arbitrary temperatures) and connected to thermal baths (with arbitrary temperatures too). Then, we will evaluate the CTP EM-field generating functional in the temporal gauge by implementing the Faddeev-Popov procedure. Moreover, we will calculate closed expressions for the EM-energy, the Poynting vector and the Maxwell tensor in terms of the Hadamard propagator, showing that all of these quantities present contributions from the field's initial conditions and also from the matter degrees of freedom in the mirrors. Then, we analyzed the dynamics of the EM field in the temporal gauge and study the case of an infinite homogeneous and isotropic material, connecting our results with previous ones. A detailed analysis is performed in relation to how the gauge invariance must be treated in the CTP formalism when the EM field is interacting with inhomogeneous anisotropic matter. We will also discuss several general implications of our results in relation to the nonequilibrium calculation of energies and forces in Casimir physics. In fact, a detailed study of the Casimir-Lifshitz problem in a fully nonequilibrium situation will be presented elsewhere using the results of this paper \cite{tumarulo}.

This paper is organized as follows: In the next Section we will discuss the CTP integration of the polarization degrees of freedom in interaction
with both, the EM-field and the thermal environment. In Section III, we evaluate the CTP-generating functional for the gauge field in the composite system. Section IV contains the formal calculation of the energy, poynting vector and mean value of the Maxwell tensor. In Section V, we give insights about the electrodynamics in the temporal gauge and analyzes a concrete example. Finally, in Section VI we will present the final remarks.

\section{CTP Integration for the Electromagnetic Field - Matter Interaction}

The main goal will be to calculate the generating functional. Therefore, the first step in this direction is to calculate the influence action over the field. We will describe the ordinary polarizable matter by (non-relativistic) quantum degrees of freedom associated to the polarization vector $\mathbf{P}$ of each volume element of the polarizable body, each one subjected to an independent bath generating an influence action as in the quantum Brownian motion (QBM) theory \cite{BreuerPett}, i. e., a quantum harmonic oscillator interacting linearly with a thermal bath consisting in a set of quantum harmonic oscillators. On the other hand, the EM field is described by a massless spin-$1$ gauge field $A^{\mu}$. The interaction term can be taken as a coupling between the field and the current generated in the polarizable matter or, equivalently, as the dipolar interaction between the polarization dipoles and the field. In other words, the interaction can be considered in two ways, depending which degree of freedom is differentiated. However, since we have non-relativistic matter degrees of freedom coupled to a gauge field which is, from the beginning, a relativistic system, differentiation is not only a time derivative, as it happens for the scalar field in Ref. \cite{RubLoLom}.

In other words, the coupling terms can be proportionals to $\int d^{4}x~P^{j}(x)~E^{j}(x)$, coupling the polarization vector to the field's canonical momentum (i. e., the electric field) in a dipole interaction way; or, analogously, $\int d^{4}x~J_{\mu}(x)~A^{\mu}(x)$, coupling the current four-vector of the (non-relativistic) polarizable matter to the (relativistic) EM field. From now on, our sum notation will be Einstein's for greek sub- and superscripts (a subscript sum with a superscript), but latin ones may sum without being opposite scripts (the covariant or contravariant nature of spatial coordinates will be adjusted by introducing minus signs when passing from a subscript to a superscript or viceversa).

Since the electric field is $E^{j}=-\partial_{j}A^{0}-\partial_{0}A^{j}$, it is clear that the interaction term is gauge invariant, while from the second way it is clear that the interaction term is a Lorentz scalar. In order to keep gauge invariance and considering also that the current $J_{\mu}$ must be a conserved current (its four-divergence must vanish), the current four-vector have to be written as $J_{\mu}=\left(\nabla\cdot\mathbf{P},-\dot{\mathbf{P}}\right)$.

After the integration over $\mathbf{P}$, we will eventually integrate over the EM field. Then, we will consider the second form for the interaction term $\int d^{4}x~J_{\mu}(x)~A^{\mu}(x)=\int d^{4}x\left(\partial_{j}P^{j}~A^{0}-\dot{P}^{j}~A^{j}\right)$ and leave a longer discussion for the next Section. However, as we will integrate over $\mathbf{P}$ firstly, we integrate by parts in the respective space coordinate for each derivative of the term involving the divergence. Therefore, considering that the field paths vanish at infinity in every direction, the interaction term reads proportional to $-\int d^{4}x\left(\partial_{j}A^{0}~P^{j}+\dot{P}^{j}~A^{j}\right)$.

The model for the total system can be described by the initial total action:

\begin{eqnarray}
S[A^{\mu},\mathbf{P}_{\mathbf{x}},\mathbf{q}_{n,\mathbf{x}}]&=&S_{0}[A^{\mu}]+S_{0}[\mathbf{P}_{\mathbf{x}}]+\sum_{n}S_{0}[\mathbf{q}_{n,\mathbf{x}}]\\
&+&S_{\rm Curr}[A^{\mu},\mathbf{P}_{\mathbf{x}}]+\sum_{n}S_{\rm int}[\mathbf{P}_{\mathbf{x}},\mathbf{q}_{n,\mathbf{x}}],\nonumber
\end{eqnarray}

\noindent where $S_{0}[A^{\mu}],S_{0}[\mathbf{P}_{\mathbf{x}}],S_{0}[\mathbf{q}_{n,\mathbf{x}}]$ are the free actions for the EM field, the polarization vector and the degrees of freedom of the thermal baths which affects the polarization vectors in each spatial point, respectively. The spatial labels denote the fact that the properties change with the position while the degree of freedom presents no spatial differentiation in its dynamics, i. e., the degrees of freedom are a spatial continuous set of $0+1$ fields. The last two actions are the interaction actions between the EM field and the polarization vectors (as we have discussed above) and between the polarization vectors and the thermal baths in each point (which are linear couplings).

Therefore, the first step would be CTP integration over the thermal baths ($\{\mathbf{q}_{n,\mathbf{x}}\}$). However, from the well-known QBM theory we know that this is already done since, as we mentioned before, it gives that the polarization vectors under the influence of the thermal baths behave effectively as Brownian particles (see Refs.\cite{BreuerPett,CalHu}). Then, for each $j-$component in each point of space, the polarization vector will have it unitary evolution modified by the QBM influence action:

\begin{eqnarray}
S_{\rm IF}[\mathbf{P},\mathbf{P}^{'}]&=&\int d\mathbf{x}\int_{t_{i}}^{t_{f}}dt\int_{t_{i}}^{t_{f}}dt'~\Delta P_{\mathbf{x}}^{j}(t)\Big[-2~D_{\mathbf{x}}(t,t')\nonumber\\
&&\times~\Sigma P_{\mathbf{x}}^{j}(t')+\frac{i}{2}~N_{\mathbf{x}}(t,t')~\Delta P_{\mathbf{x}}^{j}(t')\Big]
\label{InfluenceActionQBM}
\end{eqnarray}

\noindent where $D_{\mathbf{x}}$ and $N_{\mathbf{x}}$ are the QBM's dissipation and noise kernels respectively, while $\Delta P=P'-P$ and $\Sigma P=(P+P')/2$ are called difference and sum variables each one.

We are interested in calculating the CTP integral associated to the calculation of the influence action which acts over the EM field due to the interaction with matter. Considering a coupling constant $\lambda_{0}$ between the field and the polarization degrees of freedom, the integral is given by:

\begin{eqnarray}
&&e^{iS_{\rm IF}[A^{\mu},A'^{\mu}]}=\int d\mathbf{P}_{\rm f}\int d\mathbf{P}_{\rm i}~d\mathbf{P}'_{\rm i}\int_{\mathbf{P}(t_{\rm i})=\mathbf{P}_{\rm i}}^{\mathbf{P}(t_{\rm f})=\mathbf{P}_{\rm f}}\mathcal{D}\mathbf{P}\times \nonumber\\
&&\int_{\mathbf{P}'(t_{\rm i})=\mathbf{P}'_{\rm i}}^{\mathbf{P}'(t_{\rm f})=\mathbf{P}_{\rm f}}\mathcal{D}\mathbf{P}'~e^{i\lambda_{0}\int d\mathbf{x}~g(\mathbf{x})\left(\nabla A^{0}\cdot\mathbf{P}+\mathbf{A}\cdot\dot{\mathbf{P}}-\nabla A'^{0}\cdot\mathbf{P}'-\mathbf{A}'\cdot\dot{\mathbf{P}}'\right)}\nonumber\\
&&\times~e^{i\left(S_{0}[\mathbf{P}]-S_{0}[\mathbf{P}']+S_{\rm IF}[\mathbf{P},\mathbf{P}']\right)}~\rho_{\mathbf{P}}\left(\mathbf{P}_{\rm i},\mathbf{P}'_{\rm i},t_{\rm i}\right),
\label{FieldInfluenceAction}
\end{eqnarray}
where $A\cdot B\equiv\int_{t_{i}}^{t_{f}} dt~A(t)~B(t)$ and, for simplicity, products between vectors and matrices will be omitted in the vectorial form, and then for example $\mathbf{A}\cdot\mathbf{B}=\int_{t_{i}}^{t_{f}} dt~A^{j}(t)~B^{j}(t)$ and so on. The matter distribution function $g$, which takes binary values (1 or 0) whether or not there is matter in each point of space $\mathbf{x}$, is introduced to denote the fact that this calculation takes place in every point of space that contains polarizable material. Therefore, the influence action that acts over the EM field will be defined in space by the matter distribution function $g$, which will define in which points the influence action does not vanish.

As we are dealing with a three dimensional problem, the polarization vectors $\mathbf{P}$ described as three-dimensional harmonic oscillators can be decomposed in cartesian components, each one suffering the action of different baths. Therefore, we trivially have that $S_{0}[\mathbf{P}]=\sum_{j=1}^{3}S_{0}[P^{j}]$ and $S_{\rm IF}[\mathbf{P},\mathbf{P}']=\sum_{j=1}^{3}S_{\rm IF}[P^{j},P'^{j}]$.

Considering a separable initial state for the polarization vector, the initial density matrix is the product of density matrices for each component of the polarization vector, $\rho_{\mathbf{P}}\left(\mathbf{P}_{\rm i},\mathbf{P}'_{\rm i},t_{\rm i}\right)=\prod_{j=1}^{3}\rho_{P^{j}}\left(P_{\rm i}^{j},P_{\rm i}^{'j},\\t_{\rm i}\right)$. Then, finally noting that the interaction term also separates in each component, the CTP integral can be written as the product of three integrals for each component or direction. One step further can be done by taking advantage from the fact that the material is conceived as a continuous of independent degrees of freedom (there is no interactions between them), giving the chance for skeletonizing the spatial grid in volume elements $\Delta\mathbf{x}$ and writing the integrals for each component as a product over the spatial points ($\prod_{\mathbf{x}}$) of a generic CTP-integral with a spatial label $\mathbf{x}$.

We now proceed to write the actions in terms of the difference and sum variables. For this purpose, we consider that in each point of space for the interaction terms we have

\begin{equation}
\partial_{j}A^{0}\cdot P^{j}-\partial_{j}A^{'0}\cdot P^{'j}=-\partial_{j}\Sigma A^{0}\cdot\Delta P^{j}-\partial_{j}\Delta A^{0}\cdot\Sigma P^{j},
\end{equation}

\begin{eqnarray}
A^{j}\cdot\dot{P}^{j}-A^{'j}\cdot\dot{P}^{'j}&=&-\Sigma A^{j}~\Delta P^{j}\Big|_{t_{\rm i}}^{t_{\rm f}}-\Delta A^{j}~\Sigma P^{j}\Big|_{t_{\rm i}}^{t_{\rm f}}\nonumber\\
&+&\Sigma \dot{A}^{j}\cdot\Delta P^{j}+\Delta \dot{A}^{j}\cdot\Sigma P^{j},
\end{eqnarray}
and for the free actions we can write

\begin{eqnarray}
S_{0}[P^{j}]&-&S_{0}[P^{'j}]=-\int d\mathbf{x}~M_{\mathbf{x}}~\Sigma\dot{P}^{j}_{\mathbf{x}}~\Delta P^{j}_{\mathbf{x}}\Big|_{t_{\rm i}}^{t_{\rm f}}\\
&+&\int d\mathbf{x}\int_{t_{\rm i}}^{t_{\rm f}}dt~M_{\mathbf{x}}~\Delta P^{j}_{\mathbf{x}}\left(\frac{d^{2}}{dt^{2}}+\Omega^{2}_{\mathbf{x}}\right)\Sigma P^{j}_{\mathbf{x}},\nonumber
\end{eqnarray}
where we have allowed each degree of freedom, at each point $\mathbf{x}$, to have its own properties (mass and natural frequency).

The functional integrations over $\Delta P^{j}$ are Gaussians in each point of space, and can be performed straightforwardly considering the noise kernels in each direction and point, and defining the linear sources as $R^{j}(t)=\Delta\mathbf{x}\int_{t_{\rm i}}^{t_{\rm f}}dt'~\widetilde{L}(t,t')~\Sigma P^{j}(t')+\Delta\mathbf{x}~\lambda_{0,\mathbf{x}}\left(\Sigma\dot{A}^{j}(t)-\partial_{j}\Sigma A^{0}(t)\right)$, with the kernel $\widetilde{L}(t,t')\equiv M_{\mathbf{x}}\left(\frac{d^{2}}{dt'^{2}}+\Omega^{2}_{\mathbf{x}}\right)\delta(t-t')+D_{\mathbf{x}}(t,t')$.

At this point, the remaining functional integration is over $\Sigma P^{j}$ in each point of space. However, as it was done in Refs. \cite{RubLoLom} and \cite{CalRouVer}, we can write every path $\Sigma P^{j}$ in terms of an homogenous solution $P_{0}^{j}(t)$ satisfying the initial conditions and a particular solution $P_{\xi}^{j}(t)$ describing the deviation of the paths from the homogeneous ones, i. e., we write $\Sigma P^{j}(t)=P_{0}^{j}(t)+P_{\xi}^{j}(t)$. Considering that, from the initial actions, the canonical momentum associated to $P^{j}$ is given by $M\dot{P}^{j}+\lambda_{0}A^{j}$, the solutions can be written as

\begin{eqnarray}
P_{0}^{j}(t)&=&M_{\mathbf{x}}~\Sigma P_{\rm i}^{j}~\dot{G}_{\rm Ret,\mathbf{x}}(t-t_{\rm i})\\
&&+\left(\Pi_{\rm i}^{j}-\lambda_{0,\mathbf{x}}~\Sigma A_{\rm i}^{j}\right)G_{\rm Ret,\mathbf{x}}(t-t_{\rm i}),\nonumber
\label{P0J}
\end{eqnarray}

\begin{equation}
P_{\xi}^{j}(t)=\int_{t_{i}}^{t}ds~G_{\rm Ret,\mathbf{x}}(t-s)~\xi^{j}(s).
\end{equation}
with $\Pi_{\rm i}^{j}=M~\Sigma\dot{P}_{\rm i}^{j}+\lambda_{0}~\Sigma A_{\rm i}^{j}$ and being $G_{\rm Ret,\mathbf{x}}(t-t')$ the retarded Green function for the linear integro-differential operator, associated to the kernel $\widetilde{L}$ of the degree of freedom at $\mathbf{x}$.

Therefore, we can replace the functional integration over $\Sigma P^{j}$ with integration limits, by the functional integration over $\xi$ without them, and an ordinary integration over all the values of the initial canonical momentum $\Pi_{\rm i}^{j}$.

Then, after realizing the replacements it turns out that the functional integration over $\xi$ is straightforward and, by omitting for simplicity the spatial labels, we obtain for each integral in each point of space:

\begin{eqnarray}
&&e^{iS_{\rm IF}[A^{\mu},A'^{\mu}]}=\prod_{j,\mathbf{x}}\int d\Sigma P_{\rm i}^{j}~d\Sigma P_{\rm f}^{j}\int d\Delta P_{\rm f}^{j}\delta(\Delta P_{\rm f}^{j})d\Pi_{\rm i}^{j}\times \nonumber\\
&& e^{i\Delta\mathbf{x}\lambda_{0}\left(\Delta\dot{A}^{j}-\partial_{j}\Delta A^{0}\right)\cdot P_{0}^{j}} W_{P^{j}}\left(\Sigma P_{\rm i}^{j},\Pi_{\rm i}^{j},t_{\rm i}\right)e^{i\Delta\mathbf{x}\lambda_{0}\Delta A_{\rm i}^{j}\Sigma P_{\rm i}^{j}}\nonumber\\
&&\times~e^{-i\Delta\mathbf{x}\left(M~\Sigma\dot{P}_{\rm f}^{j}+\lambda_{0}~\Sigma A_{\rm f}^{j}\right)\Delta P_{\rm f}^{j}}~\delta\left(\Sigma P^{j}(t_{\rm f})-\Sigma P_{\rm f}^{j}\right)\nonumber\\
&&\times~e^{-\Delta\mathbf{x}\frac{\lambda_{0}^{2}}{2}\left(\Delta\dot{A}^{j}-\partial_{j}\Delta A^{0}\right)\cdot G_{\rm Ret}\cdot N\cdot\left[\left(\Delta\dot{A}^{j}-\partial_{j}\Delta A^{0}\right)\cdot G_{\rm Ret}\right]^{T}}\nonumber\\
&&\times~e^{i\Delta\mathbf{x}~\lambda_{0}^{2}\left(\Delta\dot{A}^{j}-\partial_{j}\Delta A^{0}\right)\cdot G_{\rm Ret}\cdot\left(\partial_{j}\Sigma A^{0}-\Sigma\dot{A}^{j}\right)},
\label{IntermedioInfluenceFunctional}
\end{eqnarray}
where we have introduced a delta function in order to take into account the restriction on the final points, and the Wigner function for the $j$-component of the polarization vector in the point $\mathbf{x}$ defined as in Refs. \cite{RubLoLom} and \cite{CalRouVer} by

\begin{equation}
W_{x}(X,p,t)=\frac{1}{2\pi}\int_{-\infty}^{+\infty}d\Delta e^{i\Delta\mathbf{x} p\Delta} \rho_{x}\left(X-\frac{\Delta}{2},X+\frac{\Delta}{2},t\right).
\label{WignerFunction}
\end{equation}

Now, on one hand, we note that the homogeneous solution can be written as $P_{0}^{j}(t)=P_{0}^{S,j}(t)-\lambda_{0,\mathbf{x}}~\Sigma A_{\rm i}^{j}~G_{\rm Ret,\mathbf{x}}(t-t_{i})$, where $P_{0}^{S,j}$ is the homogeneous solution for a bi-lineal coupling (where the canonical momentum is only related to the time derivatives of the degree of freedom, see for example Ref.\cite{RubLoLom}). On the other hand, we can take the exponent of the last factor and integrate by parts on the second time integration, obtaining

\begin{eqnarray}
G_{\rm Ret,\mathbf{x}}\cdot\Sigma\dot{A}^{j}&=&\int_{t_{\rm i}}^{t_{\rm f}}dt'~G_{\rm Ret,\mathbf{x}}(t-t')~\Sigma\dot{A}^{j}(t')\\
&=&-G_{\rm Ret,\mathbf{x}}(t-t_{i})~\Sigma A_{\rm i}^{j}-\partial_{t'}G_{\rm Ret,\mathbf{x}}\cdot\Sigma A^{j},\nonumber
\end{eqnarray}
where we have used that $G_{\rm Ret,\mathbf{x}}(t-t_{\rm f})=0$ because it is a retarded Green function involving a Heaviside function.

By replacing into the CTP integral, we find that one factor constituted by the second term of the last expression of $P_{0}^{j}$ in Eq.(\ref{P0J}) cancels with a factor constituted by the first term of the right hand side of last equation.

As a last step to cancel out the factors involving initial conditions, we must consider the first term in the exponent of the first factor in Eq.(\ref{IntermedioInfluenceFunctional}). Again, by integrating by parts and using the CTP condition for the EM field, we have

\begin{equation}
\Delta\dot{A}^{j}\cdot P_{0}^{S,j}=-\Delta A_{\rm i}^{j}~\Sigma P_{\rm i}^{j}-\Delta A^{j}\cdot\dot{P}_{0}^{S,j}.
\end{equation}

Then, the first term in this expression cancels out with another factor containing initial conditions.

Finally, it is straightforward to integrate over $\Delta P_{\rm f}^{j}$ by using the delta function. After this step, it is easy to evaluate the integral over $\Sigma P_{\rm f}^{j}$ by using the other delta function. Considering that the result is for the $j$-component of the polarization vector at the point $\mathbf{x}$, then, we obtain the product over all the positions where there is material. Grouping each factor and taking the continuum limit for the spatial grid, the exponents result in integrals limited by the matter distribution $g(\mathbf{x})$, thus

\begin{eqnarray}
&&e^{iS_{\rm IF}[A^{\mu},A'^{\mu}]}\\
&&=\prod_{j}\Big\langle e^{-i\int d\mathbf{x}~g(\mathbf{x})~\lambda_{0,\mathbf{x}}\left(\Delta A^{j}\cdot\dot{P}_{0}^{S,j}+\partial_{j}\Delta A^{0}\cdot P_{0}^{S,j}\right)}\Big\rangle_{\Sigma P_{\rm i}^{j},\Pi_{\rm i}^{j}}\nonumber\\
&&e^{-\frac{1}{2}\int d\mathbf{x}g(\mathbf{x})\Delta\left(\dot{A}^{j}-\partial_{j}A^{0}\right)\cdot\lambda_{0,\mathbf{x}}^{2} 
 G_{\rm Ret,\mathbf{x}}\cdot N_{\mathbf{x}}\cdot\left[\Delta\left(\dot{A}^{j}-\partial_{j}A^{0}\right)\cdot G_{\rm Ret,\mathbf{x}}\right]^{T}}\nonumber\\
&& e^{i\int d\mathbf{x}~g(\mathbf{x})\lambda_{0,\mathbf{x}}^{2}\left(\Delta\dot{A}^{j}-\partial_{j}\Delta A^{0}\right)\cdot\left(\partial_{t'}G_{\rm Ret,\mathbf{x}}\cdot\Sigma A^{j}+G_{\rm Ret,\mathbf{x}}\cdot\partial_{j}\Sigma A^{0}\right)},\nonumber 
\end{eqnarray}
where $\langle...\rangle_{\Sigma P_{\rm i}^{j},\Pi_{\rm i}^{j}}=\prod_{\mathbf{x}}\int d\Sigma P_{\rm i,\mathbf{x}}^{j}\int d\Pi_{\rm i,\mathbf{x}}^{j}~...W_{P^{j}}\Big(\Sigma P_{\rm i,\mathbf{x}}^{j},\\ \Pi_{\rm i,\mathbf{x}}^{j},t_{\rm i}\Big)$ and we let each material's property (charge, noise kernel and Green function) to depends on position by introducing a subscript $\mathbf{x}$ as a label, allowing the material to be inhomogeneous.

Now, we have to consider that this last expression is for the $j$-component of the polarization vector. The final expression for the field's influence action on Eq. (\ref{FieldInfluenceAction}) results from the product of this expression for each component. At this point, we can give the material another freedom, which is to be anisotropic (birefringent material). Each direction $j$ of the polarization vector can be matched with each of the three principal axes of Fresnel's (or refractive index) ellipsoid in each point of the material (we give more insights of this in Sec.\ref{EMFEITTG}). Therefore, each component can have different properties (except for the charge), and by introducing delta functions in space in the last two factors, we can write, more compactly

\begin{eqnarray}
&&e^{iS_{\rm IF}[A^{\mu},A^{'\mu}]}=\\
&&=\Big\langle e^{-i\int d\mathbf{x}~g(\mathbf{x})~\lambda_{0,\mathbf{x}}\left(\Delta \mathbf{A}\cdot\dot{\mathbf{P}}_{0}^{S,[j]}+\nabla\Delta A^{0}\cdot\mathbf{P}_{0}^{S,[j]}\right)}\Big\rangle_{\Sigma \mathbf{P}_{i},\mathbf{\Pi}_{i}}\nonumber\\
&&\times~e^{-\frac{1}{2}\left(\Delta\dot{\mathbf{A}}-\nabla\Delta A^{0}\right)\ast\mathbb{N}_{B}\ast\left(\Delta\dot{\mathbf{A}}-\nabla\Delta A^{0}\right)}\nonumber\\
&&\times~e^{i2\left(\Delta\dot{\mathbf{A}}-\nabla\Delta A^{0}\right)\ast\left(\partial_{t'}\mathbb{D}\ast\Sigma \mathbf{A}+\mathbb{D}\ast\nabla\Sigma A^{0}\right)},\nonumber
\end{eqnarray}
where the product $A\ast B\equiv\int d^{4}x~A(\mathbf{x},t)~B(\mathbf{x},t)$, while:

\begin{equation}
\mathbb{D}_{jk}(x,x')=\delta_{jk}~\delta(\mathbf{x}-\mathbf{x}')~g(\mathbf{x})~\frac{\lambda_{0,\mathbf{x}}^{2}}{2}~G_{\rm Ret,\mathbf{x}}^{[j]}(t-t'),
\end{equation}

\begin{equation}
\mathbb{N}^{B}_{jk}(x,x')=\delta_{jk}~\delta(\mathbf{x}-\mathbf{x}')~g(\mathbf{x})~\lambda_{0,\mathbf{x}}^{2}~G_{\rm Ret,\mathbf{x}}^{[j]}\cdot N_{\mathbf{x}}^{[j]}\cdot\left[G_{\rm Ret,\mathbf{x}}^{[j]}\right]^{T},
\label{NoiseKernelMathbbB}
\end{equation}

\noindent which are the bi-linear dissipation matrix-kernel associated to the EM field - matter interaction model (see Ref. \cite{RubLoLom}) and the bilinear noise matrix-kernel which is only related to the contribution of the baths. The superscripts $[j]$ denote the dependence on direction (anisotropic material).

To give a closed expression for the influence action on the field, we have firstly to move the temporal and spatial derivatives on the field's components to the bi-linear matrix-kernels to define correctly the `current' matrix-kernels which act over the EM field. This is in fact straightforward due to the causal behavior of the Green function, the CTP condition on the EM field's components and the convergence of each path of the EM field whenever any spatial coordinate goes to infinity. These allow us to integrate by parts both in time and spatial coordinates resulting in a transfer of the derivative on the field to the matrix-kernels. Then, we can write a covariant form for the last two factors

\begin{eqnarray}
&&e^{iS_{\rm IF}[A^{\mu},A^{'\mu}]}=\\
&&=\Big\langle e^{-i\int d\mathbf{x}~g(\mathbf{x})~\lambda_{0,\mathbf{x}}\left(\Delta \mathbf{A}\cdot\dot{\mathbf{P}}_{0}^{S,[j]}+\nabla\Delta A^{0}\cdot\mathbf{P}_{0}^{S,[j]}\right)}\Big\rangle_{\Sigma \mathbf{P}_{\rm i},\mathbf{\Pi}_{\rm i}}\nonumber\\
&&\times~e^{-\frac{1}{2}~\Delta A^{\mu}\ast N_{\mu\nu}^{B}\ast\Delta A^{\nu}}~e^{-i2\Delta A^{\mu}\ast D_{\mu\nu}\ast\Sigma A^{\nu}},\nonumber
\label{InfluenceActionFieldGeneralInitialState}
\end{eqnarray}
with the (covariant) EM dissipation kernels $D_{\mu\nu}$ and the (also covariant) EM noise kernel $N_{\mu\nu}^{B}$ associated to the contribution of the baths are given by

\begin{equation}
D_{\mu\nu}(x,x')=\Gamma_{\mu\nu}\,^{jk}~\mathbb{D}_{jk},
\label{EMDissipationKernel}
\end{equation}

\begin{equation}
N_{\mu\nu}^{B}(x,x')=\Gamma_{\mu\nu}\,^{jk}~\mathbb{N}_{jk}^{B},
\label{EMNoiseKernelBath}
\end{equation}
where the operator $\Gamma_{\mu\nu}\,^{jk}\equiv\delta_{\mu}\,^{0}~\delta_{\nu}\,^{0}~\partial_{jk'}^{2}-\delta_{\mu}\,^{0}~\delta_{\nu}\,^{k}~\partial_{jt'}^{2}-\delta_{\mu}\,^{j}~\delta_{\nu}\,^{0}~\partial_{tk'}^{2}+\delta_{\mu}\,^{j}~\delta_{\nu}\,^{k}~\partial_{tt'}^{2}$ with the prime denoting derivation on the respective coordinate of the point $x'$ and the covariant delta is introduced with Einstein's notation, unlike all delta in matrix notation employed at the moment for spatial sub and superscripts.

It is clear that the first factor on the right hand of Eq.(\ref{InfluenceActionFieldGeneralInitialState}) is entirely related to the initial state of the polarization degrees of freedom. However, in order to obtain an expression for the field's influence functional, we have to calculate the factor for the chosen initial state. This can be easily done for the case that the initial state is a thermal one for each direction of the polarization in each volume element. Therefore, considering temperatures $\beta_{P^{j}_{\mathbf{x}}}$ for each direction in each point, the integrals over $\Sigma P^{j}_{\rm i}$ and $\Pi^{j}_{\rm i}$ in each point of space are Gaussian, obtaining by discarding the normalization factor:

\begin{eqnarray}
&&\Big\langle e^{-i\int d\mathbf{x}~g(\mathbf{x})~\lambda_{0,\mathbf{x}}\left(\Delta \mathbf{A}\cdot\dot{\mathbf{P}}_{0}^{S,[j]}+\nabla\Delta A^{0}\cdot\mathbf{P}_{0}^{S,[j]}\right)}\Big\rangle_{\Sigma \mathbf{P}_{\rm i},\mathbf{\Pi}_{\rm i}}=\nonumber\\
&&=e^{-\frac{1}{2}~\Delta A^{\mu}\ast N_{\mu\nu}^{P}\ast\Delta A^{\nu}},
\end{eqnarray}
where the EM noise kernel associated to the polarization degrees of freedom is given by

\begin{equation}
N_{\mu\nu}^{P}(x,x')=\Gamma_{\mu\nu}\,^{jk}~\mathbb{N}_{jk}^{P},
\label{EMNoiseKernelPol}
\end{equation}
with:

\begin{eqnarray}
\mathbb{N}_{jk}^{P}(x,x')&=&\delta_{jk}~\delta(\mathbf{x}-\mathbf{x}')~\frac{g(\mathbf{x})\lambda_{0,\mathbf{x}}^{2}M_{\mathbf{x}}^{[j]}}{2\Omega^{[j]}_{\mathbf{x}}}\coth\left(\frac{\beta_{P^{j}_{\mathbf{x}}}~\Omega^{[j]}_{\mathbf{x}}}{2}\right)\nonumber\\
&\times&\left[\dot{G}^{[j]}_{\rm Ret,\mathbf{x}}(t-t_{i})~\dot{G}^{[j]}_{\rm Ret,\mathbf{x}}(t'-t_{i})\right.\nonumber\\
&&\left.+~\Omega^{[j]2}_{\mathbf{x}}~G^{[j]}_{\rm Ret,\mathbf{x}}(t-t_{i})~G^{[j]}_{\rm Ret,\mathbf{x}}(t'-t_{i})\right].
\label{NoiseKernelMathbbP}
\end{eqnarray}

Therefore, the EM field's influence action reads as

\begin{eqnarray}
S_{\rm IF}[A^{\mu},A'^{\mu}]&=&\int d^{4}x\int d^{4}x'~\Delta A^{\mu}(x)\Big[-2~D_{\mu\nu}(x,x')\nonumber\\
&\times&\Sigma A^{\nu}(x')+\frac{i}{2}~N_{\mu\nu}(x,x')~\Delta A^{\nu}(x')\Big],
\label{EMFieldInfluenceFunctional}
\end{eqnarray}
with $N_{\mu\nu}\equiv N_{\mu\nu}^{P}+N_{\mu\nu}^{B}$, which satisfies $N_{\mu\nu}(x,x')=N_{\nu\mu}\\(x',x)$.

First, it is worth noting that it has the form of Eq.(\ref{InfluenceActionQBM}). This is in principle not surprising, however the present influence action for the EM field is the result of two CTP integrations since the system is a composite of three parts, and the fact that it has this closed form is a merit of the choice of a thermal initial state for the polarization degrees of freedom. For a non-thermal initial state, this form is not achieved. Moreover, while the kernels in the QBM theory depend on the difference of the arguments, the present EM kernels do not.

It is remarkable that this last expression is analogous to the one found for the scalar field case for the polarization degrees of freedom contribution in Ref. \cite{RubLoLom}. However, since in this case we are dealing with an abelian gauge field, all the kernels contain the differential operator $\Gamma_{\mu\nu}^{~~~jk}$, which basically ensures that the influence action is gauge invariant. In fact, it is easy to see that $\partial^{\mu}\Gamma_{\mu\nu}^{~~~jk}=\partial^{'\nu}\Gamma_{\mu\nu}^{~~~jk}\equiv 0$, which gives that every kernels' four-divergences vanish, i. e., $\partial^{\mu}D_{\mu\nu}=\partial^{\mu}N_{\mu\nu}=\partial^{'\nu}D_{\mu\nu}=\partial^{'\nu}N_{\mu\nu}=0$. This is the property needed to ensure gauge invariance, so it is the physical requirement for every EM field's influence action of the form of Eq.(\ref{EMFieldInfluenceFunctional}) in order to be gauge invariant. In other words, every quadratic influence action must contain EM dissipation and noise kernels with null four-divergence in both subscripts.

Nevertheless, this is indeed an expected mathematical requirement which cames from the physical fact that the kernels are basically correlations functions of the current four-vector $J_{\mu}$, which is a conserved current thanks to the gauge invariance of the EM theory. If we break gauge invariance, the current is not neccesarily conserved as it happens for the Proca field theory (see Ref. \cite{Greiner}), so all the obtained properties are clearly expected as requirements to satisfy by a physically consistent EM theory.

\section{CTP Generating Functional for a Gauge Field}

In this Section we will calculate the CTP generating functional for the gauge field. Clearly, the result obtained in the last Section for the EM-field influence action will be the starting point. We are dealing with a spin-1 abelian gauge field $A^{\mu}=(A^{0},\mathbf{A})$ (being $A^{0}$ and $\mathbf{A}$ the electric and vector potentials respectively) under the influence of matter degrees of freedom, modeled through the EM dissipation and noise kernels in the influence action obtained in the last Section. Although we will use the main result in Eq.(\ref{EMFieldInfluenceFunctional}), we resume briefly some points of the previous discussion about the interaction terms.

It is worth noting that, if we consider a typical spin-0 scalar field instead of the EM field, the calculation of the generating functional can be easily done as it was shown in Ref. \cite{RubLoLom} for both bilineal and current-type couplings, without critical changes and obtaining the same formal result. In contrast, for the case of the EM field, a few subtle points must be taken into account.

Considering the gauge symmetry of the EM theory, the interaction term with matter must be gauge invariant. Thus, bilineal coupling models (as the one considered in Ref. \cite{RubLoLom}) are forbidden from the very beginning because the interaction term in the initial actions is not gauge invariant. In fact, the influence action for that
case is not gauge invariant since the kernels do not verify the correct properties that ensures gauge invariance.

However, the current-type coupling presents a subtle conceptual variation. This type of coupling cannot be on the field's time derivative as it happens for the scalar field in Ref. \cite{RubLoLom} because it also breaks gauge symmetry. Then, the interaction term must be in terms of the electric and magnetic fields in order to keep the gauge invariance of the whole theory. The general rule that is behind all of these choices is that the interaction for a current-type coupling must be on the canonical conjugate momentum. In the classical free EM field theory, the canonical momenta are defined as $\Pi^{\mu}\equiv F^{0\mu}=\delta^{\mu}_{i}~E^{i}$ (where $F^{\nu\mu}$ is the Maxwell strenght tensor).

From a formal point of view, it is well known that $\Pi^{0}\equiv 0$ and, consequently, the canonical momentum for the temporal component of $A^{\mu}$ is not well-defined, implying that a quantization procedure will not be so straightforward. However, we have no need to focus on this point at this time, and we will see how to deal with that problem in our formalism.

By considering the linearity of the interaction terms, we can write directly the CTP generating functional for the gauge field as a CTP-Feynman path integral:

\begin{eqnarray}
&&Z_{\rm CTP}[J_{\mu},J'_{\mu}]=\int dA^{\mu}_{\rm f}\int dA^{\mu}_{\rm i}~dA'^{\mu}_{\rm i}\int_{A^{\mu}(t_{\rm i})=A^{\mu}_{\rm i}}^{A^{\mu}(t_{\rm f})=A^{\mu}_{\rm f}}\mathcal{D}A^{\mu}\nonumber\\
&&\times\int_{A'^{\mu}(t_{\rm i})=A'^{\mu}_{\rm i}}^{A'^{\mu}(t_{\rm f})=A^{\mu}_{\rm f}}\mathcal{D}A'^{\mu}~e^{i\left(J_{\mu}\ast~A^{\mu}-J'_{\mu}\ast~A'^{\mu}\right)}~e^{i\left(S_{0}[A^{\mu}]-S_{0}[A'^{\mu}]\right)}\nonumber\\
&&\times~e^{iS_{\rm IF}[A^{\mu},A'^{\mu}]}~\rho_{\rm EM}(A^{\mu}_{\rm i},A'^{\mu}_{\rm i},t_{\rm i}).
\label{GFEMSinFaddeevPopov}
\end{eqnarray}

We can proceed by performing the Faddeev-Popov procedure to extract the redundant sums over paths on the same gauge class. Therefore, this includes a gauge fixing term (depending on the gauge choice defined by the gauge condition $F[A^{\mu}]=0$ which will be taken linear) and the determinant which gives the ghost terms in the Lagrangian, which, in our case, can be discarded since ghosts do not couple the EM field for every linear gauge condition due to the abelian nature of the field.

We re-write the delta-functionals as lagrangian gauge fixing terms in the exponential containing the free field actions. For this purpose, we have to continue the typical Faddeev-Popov procedure one step further. Then, we change the gauge condition on both delta functionals to $F[A^{\mu}]=C(x)$ and $F[A^{'\mu}]=C'(x)$ respectively, where $C$ and $C'$ are arbitrary functions of the coordinates (the determinant giving the ghosts do not change anyway). As the generating functional is independent of $C$ and $C'$, we can multiply by a CTP-type distribution $e^{-\frac{i}{2\alpha}\left(C\ast C-C'\ast C'\right)}$ (having an unique gauge fixing parameter $\alpha$ for both fields) and then integrate over the arbitrary functions. As we need to keep Eq. (\ref{GFEMSinFaddeevPopov}) valid, we must consider the Landau gauge for a given choice of the gauge condition, where $\alpha\rightarrow 0$, and we effectively re-obtain the delta-functionals from the gauge fixing exponentials \cite{Birrell}. Thus, we get

\begin{eqnarray}
&&Z_{\rm CTP}[J_{\mu},J'_{\mu}]=\lim_{\alpha\rightarrow 0}\int dA^{\mu}_{\alpha,\rm f}\int dA^{\mu}_{\alpha,\rm i}~dA'^{\mu}_{\alpha,\rm i}\\
&&\times\int_{A^{\mu}(t_{\rm i})=A^{\mu}_{\alpha,\rm i}}^{A^{\mu}(t_{\rm f})=A^{\mu}_{\alpha,\rm f}}\mathcal{D}A^{\mu}\int_{A'^{\mu}(t_{\rm i})=A'^{\mu}_{\alpha,\rm i}}^{A'^{\mu}(t_{\rm f})=A^{\mu}_{\alpha,\rm f}}\mathcal{D}A'^{\mu}~e^{i\left(J_{\mu}\ast A^{\mu}-J'_{\mu}\ast A'^{\mu}\right)}\nonumber\\
&&\times~e^{i\left(\widetilde{S}_{0}[A^{\mu}]-\widetilde{S}_{0}[A'^{\mu}]+S_{\rm IF}[A^{\mu},A'^{\mu}]\right)}~\rho_{\rm EM}(A^{\mu}_{\alpha,\rm i},A'^{\mu}_{\alpha,\rm i},t_{\rm i}),\nonumber
\end{eqnarray}
with $\widetilde{S}_{0}[A^{\mu}]=S_{0}[A^{\mu}]-\frac{1}{2\alpha}~F[A^{\mu}]\ast F[A^{\mu}]$ being the new action for the EM, field containing the typical gauge fixing term, which breaks gauge invariance for every $\alpha$, but that allow us to treat the component of the field as independent variables. It is clear that it turns out to be a crucial point that the gauge fixing parameter is unique, which is indeed very natural because both fields in the CTP formalism are needed to describe an unique quantum field, so the effective action must depends only on one parameter. On the other hand, it can be said that $\alpha$ must be unique since a CTP integral can always be written in terms of an unique CTP field, by parametrizing both branches of the CTP contour. Therefore, applying Faddeev-Popov procedure at this point, it is clear that $\alpha$ must be the same for both branches.

It is worth noting that we have properly included subscripts $\alpha$ as a reminder that when we take the Landau gauge (taking the limit $\alpha\rightarrow 0$), the initial and final field configurations must be written satisfying the gauge condition, since in that limit, the result must be the same as the one obtained by evaluating the deltas on the gauge condition at the very beginning. Indeed, we have to take care of this only for the initial and final points, and we do not have to introduce subscripts in the functional integrations since, as we will show below, the correct result will be naturally ensured by taking the limit on the Green function associated to the effective CTP action including the gauge fixing terms. The limit $\alpha\rightarrow 0$ or Landau gauge imposes the vanishing gauge condition (it sets the arbitrary function $C$), i. e., taking the limit will imply that, for each point, the field satisfies the vanishing gauge condition.

Now, we change variables to $\Delta A^{\mu}=A^{'\mu}-A^{\mu}$ and $\Sigma A^{\mu}=(A^{\mu}+A^{'\mu})/2$, what allow us to treat each component as an independent variable thanks to gauge fixing terms which break gauge invariance.
To continue the calculation, we first have to integrate by parts the free field actions, including the gauge fixing term. Therefore, we have to choose a gauge condition.

At this point, as we are considering linear gauge conditions, if the gauge condition actually contains or not a derivative on the field will generate crucial differences. If we take the temporal or axial gauge, the gauge condition can be written in general for both cases as $F[A^{\mu}]=t_{\mu}~A^{\mu}$, allowing $t_{\mu}$ to be a temporal or spatial four-vector for each case. For now, we will keep generality on this four-vector without choosing an specific gauge condition. Thus, in this case we do not have to integrate by parts because we directly have:

\begin{eqnarray}
&&-\frac{1}{2\alpha}\left(F[A^{\mu}]\ast F[A^{\mu}]-F[A^{'\mu}]\ast F[A^{'\mu}]\right)=\\
&&=\frac{1}{\alpha}F[\Delta A^{\mu}]\ast F[\Sigma A^{\mu}]=\frac{1}{\alpha}~\Delta A^{\nu}\ast t_{\nu}~t_{\mu}~\Sigma A^{\nu}.\nonumber
\end{eqnarray}

For the free field actions, we get

\begin{eqnarray}
&&\widetilde{S}_{0}[A^{\mu}]-\widetilde{S}_{0}[A^{'\mu}]=\int d\mathbf{x}~\Delta A^{\mu}~\eta_{\mu\nu}~\Sigma F^{0\nu}\Big|_{t_{\rm i}}^{t_{\rm f}}\\
&&-\int d^{4}x~\Delta A^{\mu}\left(\eta_{\mu\nu}~\partial_{\sigma}\partial^{\sigma}-\partial_{\mu}\partial_{\nu}-\frac{1}{\alpha}~t_{\mu}t_{\nu}\right)\Sigma A^{\nu}.\nonumber
\end{eqnarray}

Finally, considering that $J_{\mu}\ast~A^{\mu}-J'_{\mu}\ast~A'^{\mu}=-\Sigma J_{\mu}\ast\Delta A^{\mu}-\Delta J_{\mu}\ast\Sigma A^{\mu}$, the functional integration over $\Delta A^{\mu}$ can be easily done by taking $R_{\mu}=\int d^{4}x'~\mathcal{L}_{\mu\nu}(x,x')~\Sigma A^{\nu}(x')-\Sigma J_{\mu}(x)=\mathcal{L}_{\mu\nu}\ast\Sigma A^{\nu}-\Sigma J_{\mu}$, in analogy to the last Section, with the operator $\mathcal{L}_{\mu\nu}(x,x')\equiv\left(-\eta_{\mu\nu}~\square'+\partial'_{\mu}\partial'_{\nu}+\frac{1}{\alpha}~t_{\mu}t_{\nu}\right)\\ \delta(x-x')-2D_{\mu\nu}(x,x')$. This way, by defining the Wigner functional for the EM field as a natural (gauge dependent) extension of Eq.(\ref{WignerFunction}) (see Ref.\cite{MrowMull}), we obtain:

\begin{eqnarray}
&&Z_{\rm CTP}[\Sigma J_{\mu},\Delta J_{\mu}]=\lim_{\alpha\rightarrow 0}\int d\Sigma A_{\alpha,\rm i}^{\mu}~d\Sigma A_{\alpha,\rm f}^{\mu}\int d\Delta A_{\alpha,\rm f}^{\mu} \times \nonumber\\
&&\delta\left(\Delta A^{\mu}_{\alpha,\rm f}\right)\int_{\Sigma A^{\mu}_{\alpha,\rm i}}^{\Sigma A^{\mu}_{\alpha,\rm f}}\mathcal{D}\Sigma A^{\mu} e^{-i\Delta J_{\mu}\ast\Sigma A^{\mu}} e^{i\int d\mathbf{x} \Delta A^{\mu}_{\alpha,\rm f}\eta_{\mu\nu}\Sigma F^{0\nu}_{\alpha,\rm f}}\nonumber\\
&&\times~W_{\rm EM}\left[\Sigma A^{\mu}_{\alpha,\rm i},-\eta_{\mu\nu}\Sigma F^{0\nu}_{\alpha,\rm i},t_{\rm i}\right]~e^{-\frac{1}{2}R_{\mu}\ast(N^{-1})^{\mu\nu}\ast R_{\nu}}.
\end{eqnarray}

The next step is to write the paths of $\Sigma A^{\mu}$ in terms of it initial conditions and the retarded Green function associated to the operator $\mathcal{L}_{\mu\nu}$. Given the EM field equation of motion associated to the CTP effective action by:

\begin{eqnarray}
\Big(\eta_{\mu\nu}~\square-\partial_{\mu}\partial_{\nu}&-&\frac{1}{\alpha}~t_{\mu}t_{\nu}\Big)A^{\nu}(x)\\
&+&2\int d^{4}x'~D_{\mu\nu}(x,x')~A^{\nu}(x')=0,\nonumber
\label{EqMotionAmuAlpha}
\end{eqnarray}

\noindent then, the retarded Green function for $t>t'$ is defined by:

\begin{eqnarray}
\Big(\eta_{\mu\nu}~\square&-&\partial_{\mu}\partial_{\nu}-\frac{1}{\alpha}~t_{\mu}t_{\nu}\Big)\mathcal{G}_{\rm Ret,\alpha}^{\nu\lambda}(\mathbf{x},\mathbf{x}',t-t')\\
&+&2\int d^{4}x''~D_{\mu\nu}(x,x'')~\mathcal{G}_{\rm Ret,\alpha}^{\nu\lambda}(\mathbf{x}'',\mathbf{x}',t''-t')=0,\nonumber
\label{GreenFunctionAmu}
\end{eqnarray}
subjected to the initial conditions $\mathcal{G}_{\rm Ret,\alpha}^{\nu\lambda}(\mathbf{x},\mathbf{x}',0)=0$ and $\dot{\mathcal{G}}_{\rm Ret,\alpha}^{\nu\lambda}(\mathbf{x},\mathbf{x}',0)=\eta^{\nu\lambda}~\delta(\mathbf{x}-\mathbf{x}')$.

It is worth noting that we have remarked the fact that the retarded Green function depends on the gauge parameter $\alpha$ and on the time difference.

Thus, the solutions for each component of $\Sigma A^{\mu}$, splitting in homogeneous $A_{0,\alpha}^{\mu}$ and inhomogeneous $A_{\xi,\alpha}^{\mu}$ solutions, can be written as:

\begin{eqnarray}
\Sigma A^{\mu}(x)&=&A_{0,\alpha}^{\mu}(x)+A_{\xi,\alpha}^{\mu}(x)\nonumber\\
&=&\int d\mathbf{x}'~\dot{\mathcal{G}}_{\rm Ret,\alpha}^{\mu\nu}(\mathbf{x},\mathbf{x}',t-t_{\rm i})~\eta_{\nu\sigma}~\Sigma A_{\alpha,\rm i}^{\sigma}(\mathbf{x}')\nonumber\\
&&+\int d\mathbf{x}'~\mathcal{G}_{\rm Ret,\alpha}^{\mu\nu}(\mathbf{x},\mathbf{x}',t-t_{\rm i})~\eta_{\nu\sigma}~\Sigma \dot{A}_{\alpha,\rm i}^{\sigma}(\mathbf{x}')\nonumber\\
&&+\int d^{4}x'~\mathcal{G}_{\rm Ret,\alpha}^{\mu\nu}(\mathbf{x},\mathbf{x}',t-t')~\xi_{\nu}(x'),
\label{SigmaAmuTotalNOGauge}
\end{eqnarray}
where we have clearly denoted the fact that both initial conditions on the field and its time derivative, depend on the gauge parameter, while the four-function $\xi_{\nu}$ does not, since it represents a deviation of the solutions from the classical ones and, therefore, each component can be treated independently without $\alpha-$dependence.

However, the present step is quiet subtle, because it consists in the replacement of the functional integration over possible paths by two ordinary integrations over the initial field and momentum configurations (which involve the homogeneous solution) and a functional integration over $\xi^{\mu}$, which represents the shift in the path from the classical trajectory (and it includes the inhomogeneous solution).

In the present case, the replacement is not so easy since the canonical momenta are not proportional to the field components' time derivatives, as it always happens until this Section. Moreover, the choice on the gauge condition is a crucial point on defining these replacements. As we have noted in the beginning of this Section, it is well known that the canonical momentum for the temporal component $A^{0}$ is not well defined since $\Pi^{0}\equiv 0$. This is intimately related to the problematic position of the EM theory in order to be quantized, since the canonical quantization procedure cannot be developed in a straightforward way \cite{Greiner}. The same happens for a (non-CTP) path integral quantization, but the Faddeev-Popov procedure shows to be efficient in that context having no restriction on the values of the gauge parameter $\alpha$ \cite{Ramond}.

At this point, we showed that the Faddeev-Popov procedure allow us to treat the situation, but restricting the theory with the gauge fixing term to the Landau gauge. We will now see that taking this limit carefully we may impose the gauge condition in the CTP formalism.

Then, in the temporal gauge, the four-vector $t_{\mu}$ will be taken as a time-like four-vector, being our particular choice the simplest one, $t_{\mu}=(1,0,0,0)$. The gauge condition reads $F[A^{\mu}]=A^{0}$. Note that over all the possible choices of the functions $C$, to which the gauge condition will be equal, the Landau gauge implies the gauge condition for $C=0$, i. e., the field satisfies Landau gauge for the temporal gauge condition. This means that the solutions (and Green functions) obtained for an arbitrary value of the gauge parameter $\alpha$ must be taken to satisfy the Landau gauge. In this limit, as we can see from the retarded Green function equation [Eq. (\ref{GreenFunctionAmu})], $\mathcal{G}_{Ret,\alpha\rightarrow 0}^{0\nu}$ must be identical to zero.

On the other hand, from the EM theory, the canonical momentum is given by $\Pi^{\mu}\equiv F^{0\mu}=\delta^{\mu}_{~j}~E^{j}$. In the temporal gauge, the electric field is given by $E^{j}=-\partial_{0}A^{j}$, so the canonical momentum reads $\Pi^{\mu}=-\delta^{\mu}_{~j}~\dot{A}^{j}$. But in our theory with gauge fixing term (without imposing the gauge condition), the canonical momenta read as usual, $\Pi^{\mu}=\delta^{\mu}_{~j}\left(\partial^{j}A^{0}-\dot{A}^{j}\right)$. This makes that the initial time derivatives of each field component can be written as $\Sigma\dot{A}_{i}^{\sigma}=\delta^{\sigma}_{~0}~\Sigma\dot{A}_{i}^{0}+\delta^{\sigma}_{~j}\left(\partial^{j}\Sigma A_{i}^{0}-\Sigma \Pi_{i}^{j}\right)$, where the temporal component of the derivative is the only one that cannot be re-written in terms of the canonical momentum and the spatial derivative of the temporal component. Therefore, replacing this into the homogeneous solution $A_{0,\alpha}^{\mu}$, we may write it in terms of the canonical momenta and, after integrating by parts the terms associated to $\partial^{j}\Sigma A_{i}^{0}$ (and discard the boundary terms by convergence), the homogeneous solution reads:

\begin{eqnarray}
&&A_{0,\alpha}^{\mu}(x)=\int d\mathbf{x}'~\mathcal{G}_{\rm Ret,\alpha}^{\mu 0}(\mathbf{x},\mathbf{x}',t-t_{\rm i})~\Sigma\dot{A}_{\alpha,\rm i}^{0}(\mathbf{x}')\\
&+&\int d\mathbf{x}'\left(\dot{\mathcal{G}}_{\rm Ret,\alpha}^{\mu 0}(\mathbf{x},\mathbf{x}',t-t_{\rm i})-\partial'_{j}\mathcal{G}_{\rm Ret,\alpha}^{\mu j}(\mathbf{x},\mathbf{x}',t-t_{\rm i})\right)\nonumber \\
&\times & \Sigma A_{\alpha,\rm i}^{0}(\mathbf{x}')\nonumber\\
&-&\int d\mathbf{x}'~\dot{\mathcal{G}}_{\rm Ret,\alpha}^{\mu j}(\mathbf{x},\mathbf{x}',t-t_{\rm i})~\Sigma A_{\alpha,\rm i}^{j}(\mathbf{x}')\nonumber\\
&+&\int d\mathbf{x}'~\mathcal{G}_{\rm Ret,\alpha}^{\mu j}(\mathbf{x},\mathbf{x}',t-t_{\rm i})~\Sigma\Pi_{j,\alpha,\rm i}(\mathbf{x}').\nonumber
\end{eqnarray}

Then, the integration replacement can be done and then we can easily integrate on $\Delta A^{\mu}_{\alpha,\rm f}$ and then on $\Sigma A^{\mu}_{\alpha,\rm f}$ thanks to the delta-functionals, and also functionally over $\xi_{\mu}$, clearly obtaining:

\begin{eqnarray}
&Z_{\rm CTP}&[\Sigma J_{\mu},\Delta J_{\mu}]=\lim_{\alpha\rightarrow 0}\int d\Sigma A_{\alpha,\rm i}^{\mu}\int d\Sigma\Pi_{j,\alpha,\rm i} \nonumber \\
&\times & e^{-i\Delta J_{\mu}\ast A^{\mu}_{0,\alpha}}~W_{\rm EM}\left[\Sigma A^{\mu}_{\alpha,\rm i},-\Sigma\Pi_{j,\alpha,\rm i},t_{\rm i}\right]\nonumber\\
&\times & e^{-\frac{1}{2}\Delta J_{\mu}\ast\mathcal{G}_{\rm Ret,\alpha}^{\mu\nu}\ast N_{\nu\beta}\ast\left(\mathcal{G}_{\rm Ret,\alpha}^{\sigma\beta}\right)^{T}\ast\Delta J_{\sigma}}\nonumber\\
&\times & e^{-i\Delta J_{\mu}\ast\mathcal{G}_{\rm Ret,\alpha}^{\mu\nu}\ast\Sigma J_{\nu}}.
\end{eqnarray}

Finally, we take the Landau gauge, $\alpha\rightarrow 0$, which implies $\mathcal{G}_{\rm Ret,\alpha}^{\mu\nu}\rightarrow\delta^{\mu}_{~j}~\delta^{\nu}_{~k}~\mathcal{G}_{\rm Ret,LG}^{jk}$ on the retarded Green function and it also restricts the integrations because $\Sigma A^{\mu}_{\alpha,\rm i}\rightarrow\Sigma \mathbf{A}_{\rm i}$ and $\Sigma\dot{A}^{\mu}_{\alpha,\rm i}\rightarrow\Sigma\mathbf{\Pi}_{\rm i}=-\Sigma\dot{\mathbf{A}}_{\rm i}$. However, this point encodes the action of imposing the temporal gauge on the field equations at every time. Moreover, if we consider that the field is free for times before the initial one, setting $A^{0}=0$ in the field equations, result in three equations for the the vector potential $\mathbf{A}_{\rm i}$ and an additional condition (which is a residual condition resulting from the equation for $A^{0}$) given by $\nabla\cdot\mathbf{\Pi}_{\rm i}=0$. This completely defines the field, its components and canonical momenta for earlier times than the initial one, and also implies that the field has two independent components. Therefore, taking the limit $\alpha\rightarrow 0$ must includes, particularly, that the initial conditions check $A^{0}_{\rm i}=0$ and $\nabla\cdot\mathbf{\Pi}_{\rm i}=0$ (being transverse to the direction of propagation of each field mode). Then, we can naturally write:

\begin{eqnarray}
&&Z_{\rm CTP}[\Sigma \mathbf{J},\Delta \mathbf{J}]=\int d\Sigma\mathbf{A}_{\rm i}\int d\Sigma\mathbf{\Pi}_{\rm i}~e^{-i\Delta\mathbf{J}\ast \mathbf{A}_{0}}\\
&&\times~W_{\rm EM}\left[\Sigma\mathbf{A}_{\rm i},\Sigma\mathbf{\Pi}_{\rm i},t_{\rm i}\right]~e^{-\frac{1}{2}\Delta \mathbf{J}\ast\overleftrightarrow{\mathcal{G}}_{\rm Ret,LG}\ast \left(\partial_{tt'}^{2}\mathbb{N}\right)\ast\overleftrightarrow{\mathcal{G}}_{\rm Ret,LG}^{T}\ast\Delta \mathbf{J}}\nonumber\\
&&\times~e^{-i\Delta\mathbf{J}\ast\overleftrightarrow{\mathcal{G}}_{\rm Ret,LG}\ast\Sigma\mathbf{J}}\nonumber\\
&&=\Big\langle~e^{-i\Delta\mathbf{J}\ast\mathbf{A}_{0}}\Big\rangle_{\Sigma\mathbf{A}_{\rm i},\Sigma\mathbf{\Pi}_{\rm i}}e^{-\frac{1}{2}\Delta \mathbf{J}\ast\overleftrightarrow{\mathcal{G}}_{\rm Ret,LG}\ast \left(\partial_{tt'}^{2}\mathbb{N}\right)\ast\overleftrightarrow{\mathcal{G}}_{\rm Ret,LG}^{T}\ast\Delta \mathbf{J}}\nonumber\\
&&\times~e^{-i\Delta\mathbf{J}\ast\overleftrightarrow{\mathcal{G}}_{\rm Ret,LG}\ast\Sigma\mathbf{J}},\nonumber
\label{ZCTPGaugeTemporalFINAL}
\end{eqnarray}
where $\mathbf{A}_{\rm i}$ and $\mathbf{\Pi}_{\rm i}$ are taken as free initial conditions in the temporal gauge (and consequently perpendicular to the wave vector for each mode), having

\begin{eqnarray}
A_{0}^{j}(x)&=&-\int d\mathbf{x}'~\dot{\mathcal{G}}_{\rm Ret,LG}^{jk}(\mathbf{x},\mathbf{x}',t-t_{i})~\Sigma A_{\rm i}^{k}(\mathbf{x}')\nonumber\\
&&+\int d\mathbf{x}'~\mathcal{G}_{\rm Ret,LG}^{jk}(\mathbf{x},\mathbf{x}',t-t_{i})~\Sigma\Pi_{\rm i}^{k}(\mathbf{x}'),
\label{SolutionTemporalGauge}
\end{eqnarray}
which is the homogeneous solution for the field equation after imposing the temporal gauge.

It is worth noting that this is the natural and expected extension of the result obtained for the scalar field in Ref. \cite{RubLoLom}. Nevertheless, the gauge nature of the EM field causes that we have to choose a gauge condition in order to develop the calculations. In the temporal gauge, we see that the first factor in Eq. (\ref{ZCTPGaugeTemporalFINAL}), involving the initial conditions, enforces the calculation by introducing the initial conditions in the chosen gauge.

\section{Energy, Poynting Vector and Maxwell Tensor}

Once we have calculated the CTP generating functional for the EM field satisfying a given gauge condition, we can proceed with the calculation of the field correlation as functional derivatives of the generating functional. Initially, we have introduced four-vector as classical CTP-sources $J_{\mu},J'_{\mu}$. In the temporal gauge the generating functional is functionally dependent on the spatial coordinates of the CTP-sources four-vectors $J_{\mu},J'_{\mu}$, i. e., the generating functional depends on $\mathbf{J},\mathbf{J}'$. Therefore, correlation functions involving the temporal coordinate of the field, which are constructed from functional derivatives of the generating functional with respect to the temporal coordinate of the source four-vectors $J_{\mu},J'_{\mu}$, will vanish. This is clearly expected since we have chosen the temporal gauge, where $A^{0}\equiv 0$. Thus, as it is well known, the field correlation can be written in general as:

\begin{eqnarray}
\Big\langle\widehat{A}^{\mu}(x_{1})\widehat{A}^{\nu}(x_{2})\Big\rangle&=&\delta^{\mu}_{~j}~\delta^{\nu}_{~k}\Big\langle\widehat{A}^{j}(x_{1})\widehat{A}^{k}(x_{2})\Big\rangle\\
&=&\delta^{\mu}_{~j}~\delta^{\nu}_{~k}~\frac{\delta^{2}Z_{CTP}}{\delta J^{'j}(x_{1})\delta J^{k}(x_{2})}\Big|_{\mathbf{J}=\mathbf{J}'=0}.\nonumber
\end{eqnarray}

As in Ref. \cite{RubLoLom}, since the generating functional has a simple form of Eq.(\ref{ZCTPGaugeTemporalFINAL}), independently of the initial state of the field, we can easily compute its functional derivatives. Taking advantage of the symmetry properties of the noise kernel, we  obtain:

\begin{eqnarray}
\Big\langle\widehat{A}^{j}(x_{1})&&\widehat{A}^{k}(x_{2})\Big\rangle=\Big\langle A_{0}^{j}(x_{1})A_{0}^{k}(x_{2})\Big\rangle_{\Sigma\mathbf{A}_{\rm i},\Sigma\mathbf{\Pi}_{\rm i}}\nonumber\\
&&+\Big[\overleftrightarrow{\mathcal{G}}_{\rm Ret,LG}\ast \left(\partial_{tt'}^{2}\mathbb{N}\right)\ast\left(\overleftrightarrow{\mathcal{G}}_{\rm Ret,LG}\right)^{T}\Big]^{jk}(x_{1},x_{2})\nonumber\\
&&+\frac{1}{2}~\mathcal{G}_{\rm Jordan,LG}^{jk}(x_{1},x_{2}).
\label{CorrelationAj1Ak2}
\end{eqnarray}
where $\mathcal{G}_{\rm Jordan,LG}^{jk}(x_{1},x_{2})\equiv i\Big(\mathcal{G}_{\rm Ret,LG}^{kj}(x_{2},x_{1})-\mathcal{G}_{\rm Ret,LG}^{jk}(x_{1},\\x_{2})\Big)$ is the Jordan propagator \cite{CalHu} and $A_{0}^{j}$ is the homogeneous solution of Eq.(\ref{SolutionTemporalGauge}).

This correlation function corresponds to the Whightman function for the field in this open system and it is the EM field generalization of the results found in Refs. \cite{CalRouVer} and \cite{RubLoLom} for a quantum degree of freedom and a scalar field, respectively. In fact, considering that $\mathcal{G}_{\rm Ret,LG}^{jk}$ is real, is clear that the correlation is a complex quantity, with the imaginary part given by $\mathcal{G}_{\rm Jordan,LG}^{jk}$, whereas the real part is formed by the others two terms. It is possible to show, using the typical relations between the propagators, that the Hadamard propagator is given by

\begin{eqnarray}
&&\mathcal{G}_{\rm H,LG}^{jk}(x_{1},x_{2})\equiv\Big\langle A_{0}^{j}(x_{1})A_{0}^{k}(x_{2})\Big\rangle_{\Sigma\mathbf{A}_{\rm i},\Sigma\mathbf{\Pi}_{\rm i}}\\
&&+\Big[\overleftrightarrow{\mathcal{G}}_{\rm Ret,LG}\ast \left(\partial_{tt'}^{2}\mathbb{N}\right)\ast\left(\overleftrightarrow{\mathcal{G}}_{\rm Ret,LG}\right)^{T}\Big]^{jk}(x_{1},x_{2}),\nonumber
\label{HadamardPropagator}
\end{eqnarray}
where this expression holds for every initial state of the field, and it is clear that this propagator is also gauge dependent. Note that the Hadamard propagator has two separated contributions. One is associated to the material degrees of freedom represented by the noise kernel $\mathbb{N}$, which also splits into two contributions due to the composite nature of the material (polarization degrees of freedom plus bath in each point of space). The other contribution is entirely associated to the field's effective dynamics and the initial state.

With the field correlation at hand, we may calculate physical quantities of interest. We begin by giving a formal expression for the Poynting vector  \cite{Jackson,Pauli}. Considering that the real material considered is non-magnetic, we can define the Poynting vector as:

\begin{equation}
\widehat{S}^{j}(x_{1})=\frac{1}{4\pi}~\epsilon^{jkl}~\widehat{E}^{k}(x_{1})\widehat{B}^{l}(x_{1}),
\end{equation}
where $\epsilon^{jkl}$ is the Levi-Civita symbol and it is worth noting that the Poynting vector is a gauge invariant quantity, because the electric and magnetic fields are.

Following Refs. \cite{RubLoLom} and \cite{Birrell}, and also using the point splitting technique, the expectation value of the Poynting vector reads

\begin{eqnarray}
&&\Big\langle\widehat{S}^{j}(x_{1})\Big\rangle=-\frac{1}{4\pi}\lim_{x_{2}\rightarrow x_{1}}\epsilon^{jkl}\epsilon^{lmn}~\partial_{t_{1}}\partial_{m_{2}}\Big\langle\widehat{A}^{k}(x_{1})\widehat{A}^{n}(x_{2})\Big\rangle\nonumber\\
&&=\frac{1}{4\pi}\lim_{x_{2}\rightarrow x_{1}}\partial_{t_{1}}\Big(\partial_{k_{2}}\delta_{jn}-\partial_{j_{2}}\delta_{kn}\Big)\Big\langle\widehat{A}^{k}(x_{1})\widehat{A}^{n}(x_{2})\Big\rangle,
\end{eqnarray}
where $\partial_{t_{1}}$ denotes time derivative on the point $x_{1}$ and so on for the other derivatives.

It is important to consider that to use the point splitting technique, the correlation function must be a regularized quantity, in order to have finite results. Moreover, all the expectation values of interest are expected to be real quantities, as it happens with the Poynting vector. However, this seems not to be the case due to the fact that the propagator is complex.  But the coincidence limit combined with the symmetric definition of the Jordan propagator, make that the imaginary contribution vanishes at the end of the calculation. Then, the expectation value of the Poynting vector can be written only in terms of the (regularized) Hadamard propagator on Eq.(\ref{HadamardPropagator}) as follows:

\begin{eqnarray}
&&\Big\langle\widehat{S}^{j}(x_{1})\Big\rangle=\\
&&=\frac{1}{4\pi}\lim_{x_{2}\rightarrow x_{1}}\partial_{t_{1}}\Big(\partial_{k_{2}}~\delta_{jn}-\partial_{j_{2}}~\delta_{kn}\Big)~\mathcal{G}_{\rm H,LG}^{kn}(x_{1},x_{2}).\nonumber
\end{eqnarray}

This last equation gives the full time evolution of the Poynting vector, which inherits the two contribution splitting from the Hadamard propagator.

Once we have worked out the expression for the Poynting vector in terms of the Hadamard propagator, we may do so with the EM energy and the Maxwell (or stress) tensor. However, this is not so straighforward since these quantities for the EM field in real materials have not unique definitions. This is related to the freedom on the arbitrary definitions of the mechanical and EM contributions where the matter is coupled to the EM field (see for example the discussion for the classical theory given in Ref. \cite{Jackson} for simple linear isotropic media and the general approach on Ref. \cite{Pauli}) due to the non-coincidence between the displacement and electric vector fields inside a macroscopic material.

However, since the energy density and the Maxwell tensor are locally defined at each point of space, we can avoid the discussion by calculating them in vacuum regions, independently if there are matter bodies in other points of space, i. e., whether or not there are material boundaries. As in that regions there is no distinction between electric and displacement vector fields, the definitions of the energy and the Maxwell tensor are unique. Therefore, the quantum definitions for both quantities are given by \cite{Jackson,Pauli}:

\begin{equation}
\widehat{\mathcal{H}}_{\rm EM}(x_{1})=\frac{1}{8\pi}\left(\widehat{\mathbf{E}}^{2}(x_{1})+\widehat{\mathbf{B}}^{2}(x_{1})\right),
\end{equation}

\begin{eqnarray}
\widehat{T}^{jk}_{\rm Maxwell}(x_{1})&=&\frac{1}{4\pi}\Bigg[\widehat{E}^{j}(x_{1})~\widehat{E}^{k}(x_{1})+\widehat{B}^{j}(x_{1})~\widehat{B}^{k}(x_{1})\nonumber\\
&&-\frac{1}{2}~\delta_{jk}\left(\widehat{\mathbf{E}}^{2}(x_{1})+\widehat{\mathbf{B}}^{2}(x_{1})\right)\Bigg].
\end{eqnarray}

Using again the point splitting technique, we easily obtain for the expectation values of both quantities:

\begin{eqnarray}
\Big\langle\widehat{\mathcal{H}}_{\rm EM}(x_{1})\Big\rangle&=&\frac{1}{8\pi}\lim_{x_{2}\rightarrow x_{1}}\Big[\left(\partial_{t_{1}}\partial_{t_{2}}+\partial_{k_{1}}\partial_{k_{2}}\right)\delta_{lm}-\partial_{m_{1}}\partial_{l_{2}}\Big]\nonumber\\
&&\times~\mathcal{G}_{\rm H,LG}^{lm}(x_{1},x_{2}),
\end{eqnarray}

\begin{eqnarray}
\Big\langle\widehat{T}^{jk}_{\rm Maxwell}(x_{1})\Big\rangle&=&\frac{1}{4\pi}\lim_{x_{2}\rightarrow x_{1}}\Bigg[\partial_{t_{1}}\partial_{t_{2}} \delta_{jm}\delta_{ks} \nonumber \\ 
&+& \epsilon^{jlm}\epsilon^{krs}\partial_{l_{1}}\partial_{r_{2}}\nonumber\\
&-&\frac{1}{2}~\delta_{jk}\Big[\left(\partial_{t_{1}}\partial_{t_{2}}+\partial_{q_{1}}\partial_{q_{2}}\right)\delta_{ms}-\partial_{s_{1}}\partial_{m_{2}}\Big]\Bigg]\nonumber\\
&&\times~\mathcal{G}_{\rm H,LG}^{ms}(x_{1},x_{2}),
\end{eqnarray}
where it is clear that both quantities also inherit the two contribution splitting of the Hadamard propagator.

It is important to remark that this expression indeed gives the key quantities to study the Casimir force between bodies separated by vacuum regions in a fully non-equilibrium situation for a EM field, generalizing the results given for the scalar field in Ref. \cite{RubLoLom}.

As a final remark, we should note that, in a covariant formulation, these three quantities (Poynting vector, energy density and Maxwell tensor) are part of the (covariant) energy-momentum tensor for the EM field \cite{Pauli}. However, we have restricted our calculation for regions without material, avoiding the discussion for the definition of each quantity inside the material. The crucial point here is that, regardless of whether definition is considered, we can always write the expectation value in terms of the Hadamard propagator through the point splitting technique. Therefore, all the quantities will contain the contributions structure of the Hadamard propagator, given in Eq.(\ref{HadamardPropagator}). In a non-equilibrium situation, the transient dynamics of the EM field will have contributions from each part of the composite system. On the other hand, the long time regime ($t_{0}\rightarrow -\infty$) is expected to be defined at most by the baths' contributions and the field's initial state contribution, having different steady situations depending on the chosen initial state, as it was shown for the scalar field in Ref. \cite{RubLoLom}.

\section{Open Electrodynamics in the Temporal Gauge}

At this point we have fully developed the CTP formalism for the the EM field interacting with a linear, inhomogeneous and anisotropic material in a general context. In the last Section, we have written all the physical EM quantities in terms of the Hadamard propagator.

In this section, we are going to study a simple example in order to get a direct application of the developed formalism. We start by examining in a general way the EM equations of motion and then we focus on the dynamical aspects of the EM field in a infinite homogeneous and isotropic material.

\subsection{EM Field's Equation in the Temporal Gauge}\label{EMFEITTG}

The EM retarded Green tensor is defined by the equations of motion (Eq.(\ref{EqMotionAmuAlpha})) obtained from the EM CTP action after imposing the temporal gauge ($A^{0}\equiv 0$). For this purpose, it is important to note that these equations include the gauge fixing term for the temporal gauge, corresponding to the term containing the gauge fixing parameter $\alpha$. Considering the equation for the temporal coordinate ($\mu=0$), which is the only one which contain the gauge fixing term, we have:

\begin{equation}
\square A^{0}-\partial_{0}\partial_{\nu}A^{\nu}-\frac{1}{\alpha}A^{0}+2\int d^{4}x'~D_{0\nu}(x,x')~A^{\nu}(x')=0.
\label{EqMotionA0GaugeFixing}
\end{equation}

Now, by choosing the Landau gauge, where $\alpha\rightarrow 0$, it naturally implies that $A^{0}\equiv 0$ in order to do not have divergent terms. Then, the (vanishing) temporal gauge is naturally introduced by the choice of the Landau gauge. The equation in this case, remains:

\begin{equation}
-\partial_{0}\partial_{m}A^{m}+2\int d^{4}x'~D_{0m}(x,x')~A^{m}(x')=0.
\label{EqMotionA0TemporalGauge}
\end{equation}

The dynamical equation for the temporal component $A^{0}$ in the vanishing gauge condition becomes a residual condition for the remaining components $A^{m}$. From the definition of the EM dissipation kernel of Eq.(\ref{EMDissipationKernel}), $D_{0m}$ can be easily calculated and the condition reads:

\begin{equation}
\partial_{m}\left[\partial_{0}A^{m}-\lambda_{0,\mathbf{x}}^{2}g(\mathbf{x})\int_{t_{i}}^{t}dt'~\dot{G}_{\rm Ret,\mathbf{x}}^{[m]}(t-t')A^{m}(\mathbf{x},t')\right]=0,
\end{equation}

\noindent where we have considered that $G_{\rm Ret,\mathbf{x}}^{[m]}(t-t')$ is a function of $t-t'$ plus the distribution $\Theta(t-t')$ in order to write it derivative and that $G_{\rm Ret,\mathbf{x}}^{[m]}(0)=0$.

By writing the first term as a integral:

\begin{equation}
\partial_{0}A^{m}(\mathbf{x},t)=-\int_{t_{i}}^{t}dt'~\partial_{t'}\Big(\delta(t'-t)\Big)~A^{m}(\mathbf{x},t'),
\end{equation}

\noindent then, as for any function $f$ we have that $\partial_{t}f(t-t')=-\partial_{t'}f(t-t')$ and the fact that the derivative of the Dirac delta function is an odd function, the condition can be written in general in its vectorial form as:

\begin{equation}
\nabla\cdot\left[\int_{t_{i}}^{t}dt'~\partial_{t}\Big(\overleftrightarrow{\varepsilon}(t-t',\mathbf{x})\Big)\cdot\mathbf{A}(\mathbf{x},t')\right]=0,
\label{ResidualCondition}
\end{equation}

\noindent where the permittivity tensor for the inhomogeneous and anisotropic material is given by:

\begin{equation}
\overleftrightarrow{\varepsilon}(t-t',\mathbf{x})_{mr}\equiv\delta_{mr}\left(\delta(t-t')+\lambda_{0,\mathbf{x}}^{2}g(\mathbf{x})G_{\rm Ret,\mathbf{x}}^{[m]}(t-t')\right).
\label{PermittivityTensor}
\end{equation}

It is worth noting that since in this basis the tensor is diagonal, it turns out that it is expressed in the Fresnel's principal axes basis. In fact, it turns out that any point of space having material we have the same Fresnel's basis.  This is why, from the beginning, we considered the same three directions of oscillation for each polarization degree of freedom. Particularly, this also holds for disjoint material bodies, although in general is clear that is possible to have bodies with different Fresnel's bases. All these cases can be considered introducing changes of basis in order to relate them. However, in order to keep simplicity in the expressions these complications will be omitted in this work. Moreover, the residual condition expressed in Eq.(\ref{ResidualCondition}) is, for the case of inhomogeneous and anisotropic material, close to conditions considered in the Literature. For example, on one hand, is close to the one considered in Ref.\cite{BehuninHu2011} (for dissipative isotropic materials) as a complementary condition for the temporal gauge, but for the case of anisotropic materials.

On the other hand, the present condition is also close to the generalized Coulomb gauge condition considered in Ref.\cite{EberleinRobaschik2006}. For the case of non-dissipative and also non-dispersive isotropic media, i. e., for the case of constant dielectric permittivity, a close condition can be obtained. First, isotropy implies that superscripts $[m]$ are omitted. Now, for an arbitrary type of bath, the well-know QBM theory gives that Laplace transform of the retarded Green function for the material is given by (see Refs.\cite{BreuerPett,LombiMazziRL,RubLoLom}):

\begin{equation}
G_{\rm Ret,\mathbf{x}}(s)=\frac{1}{\Big(s^{2}+\Omega_{\mathbf{x}}^{2}-2~D_{\mathbf{x}}(s)\Big)}.
\label{QBMRetardedGreenFunction}
\end{equation}

Then, the constant dielectric permittivity case is given by setting $s=0$ in the Laplace transform (Ref.\cite{RubLoLom}):

\begin{equation}
G_{\rm Ret,\mathbf{x}}(s)\rightarrow G_{\rm Ret,\mathbf{x}}(0)=\frac{1}{\Omega_{\mathbf{x}}^{2}}\equiv G_{\rm Ret,\mathbf{x}}^{ND}.
\end{equation}

By Mellin's transform, the associated retarded Green function results:

\begin{equation}
G_{\rm Ret,\mathbf{x}}^{ND}(t-t')=\int_{\alpha-i\infty}^{\alpha+i\infty}\frac{ds}{2\pi i}~e^{s(t-t')}~G_{\rm Ret,\mathbf{x}}^{ND}=\frac{1}{\Omega_{\mathbf{x}}^{2}}~\delta(t-t').
\end{equation}

All in all, the residual condition in Eq.(\ref{ResidualCondition}) easily reads:

\begin{equation}
\nabla\cdot\left[\varepsilon(\mathbf{x})~\mathbf{E}(\mathbf{x},t)\right]=0,
\label{EberleinRobaschikCondition}
\end{equation}

\noindent where the inhomogeneous constant permittivity function results $\varepsilon(\mathbf{x})=1+\frac{\lambda_{0,\mathbf{x}}^{2}}{\Omega_{\mathbf{x}}^{2}}~g(\mathbf{x})$, which results clearly ensured by taking the generalized Coulomb gauge condition of Ref.\cite{EberleinRobaschik2006} ($\nabla\cdot\left[\varepsilon(\mathbf{x})~\mathbf{A}(\mathbf{x},t)\right]=0$) for the case of inhomogeneous materials, since in the temporal gauge $\mathbf{E}=-\partial_{0}\mathbf{A}$. Is clear that in this case, without dissipation, the permittivity function in the complex $s-$plane is real inside the material. Therefore, the refractive index $n_{\mathbf{x}}=\sqrt{\varepsilon(\mathbf{x})}$ is real in each point. Due to isotropy, the Fresnel ellipsoid's picture in each point is trivial and corresponds, in every point, to a sphere because all the ellipsoid axis are equal.

However, without assuming isotropy, a similar condition for anisotropic materials can be obtained. It is straightforward that for the anisotropic case, we would obtain the same equation as Eq.(\ref{EberleinRobaschikCondition}) by replacing the permittivity function $\varepsilon(\mathbf{x})$ with the inhomogeneous constant permittivity tensor $\overleftrightarrow{\varepsilon}(\mathbf{x})=\mathbb{I}\left(1+\frac{\lambda_{0,\mathbf{x}}^{2}}{\Omega_{\mathbf{x}}^{[j]2}}~g(\mathbf{x})\right)$. As in the last case, since the material has no dissipation, the permittivity tensor inside the material in the complex $s-plane$ is real. Therefore, the refractive indexes in each direction are $n^{[j]}_{\mathbf{x}}=\sqrt{\varepsilon^{[jj]}(\mathbf{x})}$, where there is no implicit sum in $[jj]$. Then, in each point, we define the typical (real) Fresnel's ellipsoid.

We can conclude that, for the temporal gauge, the equation of motion for $\mu=0$ reduces to a residual condition given, for the general case, by Eq.(\ref{ResidualCondition}). On the other hand, if we take the remaining equations of motion ($\mu=m$), by imposing the temporal gauge, we clearly have:

\begin{equation}
-\square A^{m}-\partial_{m}\partial_{l}A^{l}+2\int d^{4}x'~D_{ml}(x,x')~A^{l}(x')=0.
\end{equation}

In this case, considering the components of the EM dissipation kernel from Eq.(\ref{EMDissipationKernel}), it is straightforward that the equation reads:

\begin{eqnarray}
&-&\square A^{m}-\partial_{m}\partial_{l}A^{l}\\
&+&\lambda_{0,\mathbf{x}}^{2}~g(\mathbf{x})\int_{t_{i}}^{t}dt'~\partial_{tt'}^{2}\left(G_{\rm Ret,\mathbf{x}}^{[m]}(t-t')\right)A^{m}(\mathbf{x},t')=0.\nonumber
\end{eqnarray}

Considering twice the derivative property of a product between a function and a distribution plus the initial conditions for $G_{\rm Ret,\mathbf{x}}^{[m]}$, we finally obtain:

\begin{eqnarray}
&&\frac{\partial^{2}\mathbf{A}}{\partial t^{2}}+\nabla\times\left(\nabla\times\mathbf{A}\right)+\lambda_{0,\mathbf{x}}^{2}~g(\mathbf{x})~\mathbf{A}(\mathbf{x},t)\\
&&+~\lambda_{0,\mathbf{x}}^{2}~g(\mathbf{x})\int_{t_{i}}^{t}dt'~\ddot{\overleftrightarrow{\mathbf{G}}}_{\rm Ret,\mathbf{x}}(t-t')\cdot\mathbf{A}(\mathbf{x},t')=0,\nonumber
\label{EqMotionAmTemporalGauge}
\end{eqnarray}

\noindent where $\left(\overleftrightarrow{\mathbf{G}}_{\rm Ret,\mathbf{x}}\right)_{mk}=\delta_{mk}~G_{\rm Ret,\mathbf{x}}^{[m]}$. Again, from the fact that we could write this diagonal tensor associated to the retarded Green functions, it is clear that the basis that we have chosen is the Fresnel's principal axes basis (in general, the tensor would be non-diagonal). It is also remarkable the appearance of the third term, which constitutes a finite renormalization position-dependent mass term for the EM field as the one found in the scalar case in Ref.\cite{RubLoLom}.

In fact, Eq.(\ref{EqMotionAmTemporalGauge}), in some sense, can be considered as the electromagnetic or vectorial generalization of the equation of motion for the scalar field found in Ref.\cite{RubLoLom}, including all the properties related to dissipation and inhomogeneity and also to anisotropy, which is a property entirely related to the vectorial nature of the EM field. However, the equations are not formally the same, because for the scalar case the second term of the l.h.s. of Eq.(\ref{EqMotionAmTemporalGauge}) is a Laplacian, while in the present case there is one more term related to the divergence of the field $\mathbf{A}$.

In Ref.\cite{KnollLeonhardt1992}, a similar model for the interaction between the matter and the EM field is considered. For simplicity, an unidimensional problem is taken with an EM field in the Coulomb gauge from the start. The field equation is deduced by solving the Heisenberg equations of motion for the matter's degrees of freedom and a Laplacian is obtained in place of the second term of l.h.s. of Eq.(\ref{EqMotionAmTemporalGauge}) due to the Coulomb gauge condition. This is commented in the beginning in order to explain the context in which this toy model is inspired, but no formal treatment related to gauge invariance is given. This way, the field equation has obviously the same form as for the scalar field in Ref.\cite{RubLoLom}. However, the crucial point here is that the Coulomb gauge, unlike to what happens for the free field case, may not imply $A^{0}=0$. In fact, in that gauge, the four components of the EM field may still be coupled due to the term involving the EM dissipation kernel and the temporal component $A^{0}$ cannot be discarded as in Ref.\cite{KnollLeonhardt1992}. On the other hand, by taking the temporal gauge, Eq.(\ref{ResidualCondition}) is the residual condition that must be satisfied, which is not necessarily Coulomb condition for the case where there are variations in material (see next Section for the homogeneous case) and thus, the equation of motion is given by Eq.(\ref{EqMotionAmTemporalGauge}) differing from the one considered in Ref.\cite{KnollLeonhardt1992}. All in all, we can say that the toy model of Ref.\cite{KnollLeonhardt1992} shows not to represent one of the components of the EM field in the context commented (with spatial variations of the dielectric properties) and it is closer to the scalar field model considered in Ref.\cite{RubLoLom}, which fully coincides for the one dimensional case.

Nevertheless, the temporal gauge shows to be adequate for the interaction with matter since it decouples the components of the EM field. Indeed, in the present model, a realistic EM field (with gauge and vectorial properties) interacting with matter must satisfies Eq.(\ref{EqMotionAmTemporalGauge}) and the residual condition given in general by Eq.(\ref{ResidualCondition}).

Given the equation of motion, the `temporal-gauged' EM retarded Green tensor $\overleftrightarrow{\mathcal{G}}_{Ret}(\mathbf{x},\mathbf{x}',t)$ can be defined as (omitting the subscripts $LG$ for Landau gauge because, from here and so on, they are useless since the vanishing temporal gauge was already introduced):

\begin{eqnarray}
&0&= \frac{\partial^{2}\overleftrightarrow{\mathcal{G}}_{\rm Ret}}{\partial t^{2}}+\nabla\times\left(\nabla\times\overleftrightarrow{\mathcal{G}}_{\rm Ret}\right)+\lambda_{0,\mathbf{x}}^{2} g(\mathbf{x}) \overleftrightarrow{\mathcal{G}}_{\rm Ret}\\
&&+\lambda_{0,\mathbf{x}}^{2}g(\mathbf{x})\int_{t_{i}}^{t}dt''\ddot{\overleftrightarrow{\mathbf{G}}}_{\rm Ret,\mathbf{x}}(t-t'')\cdot\overleftrightarrow{\mathcal{G}}_{\rm Ret}(\mathbf{x},\mathbf{x}',t''-t'), \nonumber
\label{EqMotionEMRetGreenTensor}
\end{eqnarray}

\noindent which is now subjected to the initial conditions in the temporal gauge:

\begin{equation}
\mathcal{G}_{\rm Ret}^{jk}(\mathbf{x},\mathbf{x}',0)=0~~,~~\dot{\mathcal{G}}_{\rm Ret}^{jk}(\mathbf{x},\mathbf{x}',0)=-~\delta^{jk}~\delta(\mathbf{x}-\mathbf{x}').
\label{GreenTensorInitialConditions}
\end{equation}

Once we have studied the EM field's equations of motion in the temporal gauge, we can realize an immediate application to the simple problem of studying the steady state of the EM field in bulk homogeneous and isotropic material.

\subsection{Steady State of the EM field in a Bulk Homogeneous and Isotropic Material}

Having analyzed the EM field dynamics in the temporal gauge, we can straightforwardly study the steady state situation of the EM field in an infinite homogeneous and isotropic material. In this case, as we anticipated in the last section, a few simplifications overcome on the results recently obtained. On the one hand, homogeneity and isotropy implies to drop out all the label $\mathbf{x}$ and $[j]$ in the material properties. Moreover, since the material is infinite, $g\equiv 1$ for every point. The EM field's equation of motion in Eq.(\ref{EqMotionAmTemporalGauge}) and the residual condition in Eq.(\ref{ResidualCondition}) reads:

\begin{eqnarray}
\frac{\partial^{2}\mathbf{A}}{\partial t^{2}}&+&\nabla\times\left(\nabla\times\mathbf{A}\right)+\lambda_{0}^{2}~\mathbf{A}(\mathbf{x},t)\\
&+&\lambda_{0}^{2}\int_{t_{i}}^{t}dt'~\ddot{G}_{\rm Ret}(t-t')~\mathbf{A}(\mathbf{x},t')=0,\nonumber
\label{EqMotionAmTemporalGaugeBulkSS}
\end{eqnarray}

\begin{equation}
\nabla\cdot\left[\int_{t_{i}}^{t}dt'~\partial_{t}\varepsilon(t-t')~\mathbf{A}(\mathbf{x},t')\right]=0,
\end{equation}

\noindent where the permittivity tensor is proportional to the identity so the residual condition simplifies.

Moreover, since the permittivity now is independent of the position, the last condition is guaranteed if we have $\nabla\cdot\mathbf{A}=0$.

Therefore, for the case in which the bulk be (infinite) homogeneous and an isotropic material, Coulomb condition is naturally required by the general residual condition for every time in every point of space. This implies that also the EM field's equation of motion in Eq.(\ref{EqMotionAmTemporalGaugeBulkSS}) simplifies a little more and reduces to:

\begin{eqnarray}
\frac{\partial^{2}\mathbf{A}}{\partial t^{2}}&-&\nabla^{2}\mathbf{A}+\lambda_{0}^{2}~\mathbf{A}(\mathbf{x},t)\\
&+&\lambda_{0}^{2}\int_{t_{i}}^{t}dt'~\ddot{G}_{\rm Ret}(t-t')~\mathbf{A}(\mathbf{x},t')=0,\nonumber
\label{EqMotionAmTemporalGaugeBulk}
\end{eqnarray}

\noindent where the second term (discussed in the last Section where related with Ref.\cite{KnollLeonhardt1992}) is reduced to the Laplacian.

As always, we can now straightforwardly define the EM retarded Green tensor $\overleftrightarrow{\mathcal{G}}_{Ret}(\mathbf{x},\mathbf{x}',t)$ for this case as:

\begin{eqnarray}
&&\frac{\partial^{2}\overleftrightarrow{\mathcal{G}}_{\rm Ret}}{\partial t^{2}}-\nabla^{2}\overleftrightarrow{\mathcal{G}}_{\rm Ret}+\lambda_{0}^{2}~\overleftrightarrow{\mathcal{G}}_{\rm Ret}(\mathbf{x},\mathbf{x}',t-t')\\
&&+~\lambda_{0}^{2}\int_{t_{i}}^{t}dt''~\ddot{G}_{\rm Ret}(t-t'')\overleftrightarrow{\mathcal{G}}_{\rm Ret}
(\mathbf{x},\mathbf{x}',t''-t')=0.\nonumber
\label{EqMotionGREENTemporalGaugeBulk}
\end{eqnarray}

However, we can further take advantage of the translational invariance provided by the uniformity of the material when being infinite, homogeneous and isotropic. Thus, we can Fourier transform in the spatial variables and write the equation of motion for the EM field's Fourier transform $\mathbf{A}(\mathbf{k},t)$:

\begin{equation}
\frac{\partial^{2}\mathbf{A}}{\partial t^{2}}+\left(k^{2}+\lambda_{0}^{2}\right)\mathbf{A}(\mathbf{k},t)+\lambda_{0}^{2}\int_{t_{i}}^{t}dt' \ddot{G}_{\rm Ret}(t-t') \mathbf{A}(\mathbf{k},t')=0
\label{EqMotionAmTemporalGaugeBulkFOURIER}
\end{equation}

\noindent where $k=|\mathbf{k}|$, while the Coulomb condition reduces to $\mathbf{k}\cdot\mathbf{A}(\mathbf{k},t)=0$ (transverse waves). It is clear that two components of EM field's Fourier transform are independent. Then, choosing two of them and its associated equations of motion, the third equation of motion for the remaining component is automatically satisfied.

Regarding in the choice, the key point is that each component satisfies an homogeneous QBM equation for a Brownian oscillator having a frecuency $\sqrt{k^{2}+\lambda_{0}^{2}}$ and a damping kernel given by $\lambda_{0}^{2}~\ddot{G}_{\rm Ret}(t-t')$ (see Refs.\cite{BreuerPett,LombiMazziRL,CalHu}).

Moreover, due to translational invariance, the EM retarded Green tensor will depend on $\mathbf{x}-\mathbf{x}'$ and therefore its Fourier transform can be easily defined by the equation:

\begin{eqnarray}
\frac{\partial^{2}\overleftrightarrow{\mathcal{G}}_{\rm Ret}}{\partial t^{2}}&+&\left(k^{2}+\lambda_{0}^{2}\right)\overleftrightarrow{\mathcal{G}}_{\rm Ret}(\mathbf{k},t-t')\\
&+&\lambda_{0}^{2}\int_{t_{i}}^{t}dt''~\ddot{G}_{\rm Ret}(t-t'')~\overleftrightarrow{\mathcal{G}}_{\rm Ret}(\mathbf{k},t''-t')=0,\nonumber
\label{EqMotionGREENTemporalGaugeBulkFOURIER}
\end{eqnarray}

\noindent subjected this time to the `Fourier-transformed' initial conditions:

\begin{equation}
\mathcal{G}_{\rm Ret}^{jl}(\mathbf{k},0)=0~~,~~\dot{\mathcal{G}}_{\rm Ret}^{jl}(\mathbf{k},0)=-~\delta^{jl}.
\label{GreenTensorInitialConditionsFOURIER}
\end{equation}

It can be easily shown by Laplace-transforming the last equation of motion that the EM retarded Green tensor is diagonal and each non-vanishing component corresponds to the QBM retarded Green function with the respective damping kernel. Regarding of the damping kernel, causality implies that the poles of the Laplace transform of this Green function must lie on the left half of the complex plane and then, its dynamics will be vanishing in the long-time limit (see Ref.\cite{RubLoLom} for the scalar analog). Thus, $\mathcal{G}_{\rm Ret}^{jl}(\mathbf{k},t-t_{i})$ will be a tensor which goes to 0 when $t_{i}\rightarrow-\infty$.

With this analysis of the EM retarded Green tensor, we can study the time evolution of the Hadamard propagator of Eq.(\ref{HadamardPropagator}) (is clear that to study the energy, Poynting vector or Maxwell tensor in this case, we should consider it expressions in a material region, what we avoid specifically in Sec.V because the arbitrariness in their definitions). For the first term of the Hadamard propagator, by introducing the Fourier transform of the retarded Green tensor in the homogeneous solutions, we can easily write:

\begin{eqnarray}
&&\Big\langle A_{0}^{j}(x_{1})~A_{0}^{m}(x_{2})\Big\rangle_{\Sigma\mathbf{A}_{i},\Sigma\mathbf{\Pi}_{i}}=\int\frac{d\mathbf{k}_{1}}{(2\pi)^{3}}\int\frac{d\mathbf{k}_{2}}{(2\pi)^{3}}\\
&&\times~e^{-i\left(\mathbf{k}_{1}\cdot\mathbf{x}_{1}+\mathbf{k}_{2}\cdot\mathbf{x}_{2}\right)}\Big\langle A_{0}^{j}(\mathbf{k}_{1},t_{1})~A_{0}^{m}(\mathbf{k}_{2},t_{2})\Big\rangle_{\Sigma\mathbf{A}_{i},\Sigma\mathbf{\Pi}_{i}},\nonumber
\end{eqnarray}

\noindent where $A_{0}^{j}(\mathbf{k},t)$ by direct construction is given by:

\begin{eqnarray}
A_{0}^{j}(\mathbf{k},t)&=&-\dot{\mathcal{G}}_{\rm Ret}^{jl}(\mathbf{k},t-t_{i})\int d\mathbf{x}'~e^{i\mathbf{k}\cdot\mathbf{x}'}~\Sigma A_{i}^{l}(\mathbf{x}')\nonumber\\
&+&\mathcal{G}_{\rm Ret}^{jl}(\mathbf{k},t-t_{i})\int d\mathbf{x}'~e^{i\mathbf{k}\cdot\mathbf{x}'}~\Sigma\Pi_{i}^{l}(\mathbf{x}'),
\end{eqnarray}
where we explicitly write the integrals over $\mathbf{x}'$ since the notation  $\Big\langle ...\Big\rangle_{\Sigma\mathbf{A}_{i},\Sigma\mathbf{\Pi}_{i}}$ implies functional integrations over $\Sigma\mathbf{A}_{i}(\mathbf{x}),\Sigma\mathbf{\Pi}_{i}(\mathbf{x})$.

The key point here is that, beyond the chosen initial state for the EM field, for the long-time limit ($t_{i}\rightarrow-\infty$) we have that $\mathcal{G}_{\rm Ret}^{jl}(\mathbf{k},t-t_{i})\rightarrow 0$ and therefore $\Big\langle A_{0}^{j}(x_{1})~A_{0}^{m}\\(x_{2})\Big\rangle_{\Sigma\mathbf{A}_{i},\Sigma\mathbf{\Pi}_{i}}\longrightarrow 0$. In other words, the term associated to the initial conditions does not contribute to the steady regime.

On the other hand, the second term of Eq.(\ref{HadamardPropagator}), associated to the material's contribution present a simpler behaviour independent of the EM retarded Green tensor's properties. In this case, the initial time $t_{i}$ does not appears in the EM retarded Green tensors contained in the term, but it does in one of the parts of the EM noise kernel given in this gauge by $\partial_{tt'}^{2}\mathbb{N}$. As we have $\mathbb{N}=\mathbb{N}^{B}+\mathbb{N}^{P}$, when each term is given by Eqs.(\ref{NoiseKernelMathbbB}) and (\ref{NoiseKernelMathbbP}) respectively, the former does not contain $t_{i}$ while the latter does. In fact, from its definition it is easy to see that only $\mathbb{N}^{P}$ goes to 0 in the long-time limit ($t_{i}\rightarrow-\infty$). Then, $\mathbb{N}\rightarrow\mathbb{N}^{B}$ in the long-time limit, proving that the Hadamard propagator in the steady regime reduces to the baths' contribution:

\begin{equation}
\mathcal{G}_{\rm H}^{jk}(x_{1},x_{2})\longrightarrow\Big[\overleftrightarrow{\mathcal{G}}_{\rm Ret}\ast \left(\partial_{tt'}^{2}\mathbb{N}^{B}\right)\ast\left(\overleftrightarrow{\mathcal{G}}_{\rm Ret}\right)^{T}\Big]^{jk}(x_{1},x_{2}),
\end{equation}

\noindent which is an analog result of the one obtained for the scalar field in Ref.\cite{RubLoLom} but for the EM field, and is also in agreement with the steady functional approach employed in Ref.\cite{Bechler}.

This result is indeed physically expected because the dissipative dynamics of the EM field in every point of space. It can be also easily shown (see Ref.\cite{RubLoLom} for the scalar case) that if a non-dissipative material is considered, all the material dynamics should be erased by setting $\mathbb{N}\equiv 0$ and the Hadamard propagator would be defined only by the initial conditions' term, which in this case does not vanishes. Moreover, as it is mentioned in Ref.\cite{Bechler} and shown for an unidimensional scalar field in Ref.\cite{RubLoLom}, a caveat may be done around a scenario presenting vacuum regions (or at least, regions where the field has no damping dynamics), where more than one of the terms define the steady state. This will be the objective of forthcoming work \cite{tumarulo}.

\section{Conclusions and Forthcoming Work}

In this paper we have mainly extended a previous work \cite{RubLoLom} in order to calculate the CTP generating functional for the EM field in interaction with inhomogeneous anisotropic matter, through the open system framework. We have calculated a general expression for the EM field's influence action from the interaction of the field with a composite environment, consisting in quantum polarization degrees of freedom in each point of space, and connected to thermal baths (with arbitrary temperatures). Then, we have evaluated the CTP-EM-field generating functional in the temporal gauge by implementing the Faddeev-Popov procedure. Special care has been taken about how the gauge invariance must be treated in the CTP formalism when the EM field is interacting with inhomogeneous anisotropic matter.

In addition to previous works, we have also found closed expressions for the EM-energy, the Poynting vector and the Maxwell tensor, in vacuum regions, in terms of the Hadamard propagator, showing that all of these quantities present contributions from the field's initial conditions and also from the matter degrees of freedom in the material bodies.

We then study the open electrodynamics in the temporal gauge, obtaining the EM field's equation and a residual condition closely related to the gauge condition considered in Ref.\cite{EberleinRobaschik2006}. Finally, we analyzed the dynamics and steady regime of the Hadamard propagator for the case of an infinite homogeneous and isotropic material interacting with the EM field, showing that in this case, in the long-time limit, the only contribution that survives is the one associated to the bath due to the damping dynamics of the EM field in every point of space.

Nevertheless, despite on the result obtained for this last situation, a few general considerations are in order about the contributions that we found. In the transient evolution, all the contributions will be present. In the long time regime, however, we expect that the baths' and the field's initial state contributions were the only that contribute at most depending if there are (or not) vacuum regions (or regions where the field has no damping dynamics) in the problem, as it happens for the scalar field in Ref. \cite{RubLoLom}. If there is such a dependence with the initial state contribution, it means that the steady situation is non unique. This is clearly caused by the fact that the field fluctuates freely and without damping in the vacuum regions, making the contribution of the initial conditions to reach the long time regime. Following this train of thought, we have also discussed in detail several implications of our results in relation to the nonequilibrium calculation of energies and forces in Casimir physics. As a forthcoming work is pending the study of the Casimir-Lifshitz problem in a fully nonequilibrium situation \cite{tumarulo}, exploiting all the features of the present quantum approach.

\begin{acknowledgement}
We would like to thank F.D Mazzitelli for useful comments. This work is supported by CONICET, UBA, and ANPCyT, Argentina.
\end{acknowledgement}

\end{document}